\documentclass[journal,a4paper,10pt]{IEEEtran}
\usepackage[T1]{fontenc}
\usepackage[latin9]{inputenc}
\usepackage{epsfig,psfrag}
\usepackage{amsmath,amsfonts,amssymb,amsxtra,bm}
\usepackage{graphicx}
\usepackage{color}
\usepackage{cite}
\usepackage{tabularx}

\newcolumntype{L}[1]{>{\raggedright\arraybackslash}p{#1}}
\newcolumntype{C}[1]{>{\centering\arraybackslash}p{#1}}
\newcolumntype{R}[1]{>{\raggedleft\arraybackslash}p{#1}}
\newcommand{\eq}{\,=\,}
\newcommand{\be}{\begin{equation}}
\newcommand{\ee}{\end{equation}}
\newcommand{\ist}{\hspace*{.3mm}}
\newcommand{\rmv}{\hspace*{-.3mm}}

\newcommand{\rrmv}{\hspace*{-1mm}}
\newcommand{\nn}{\nonumber}
\newcommand{\bd}[1]{\mathbf{#1}}
\newcommand{\cl}[1]{\mathcal{#1}}
\newcommand{\EE}{\mathcal{E}}
\renewcommand{\AA}{\mathcal{A}}
\newcommand{\OO}{\mathcal{O}}

\allowdisplaybreaks

\begin{document}
\title{Distributed Localization and Tracking of Mobile Networks Including Noncooperative Objects --- Extended Version\vspace{2mm}}

\author{Florian Meyer, \emph{Member, IEEE}, Ondrej Hlinka, Henk Wymeersch, \emph{Member, IEEE}, Erwin Riegler, \emph{Member, IEEE}, and Franz Hlawatsch, \emph{Fellow, IEEE}
\thanks{Extended version of a manuscript accepted for publication in the IEEE Transactions on Signal and Information Processing over Networks, 2016.}
\thanks{F.\ Meyer and F.\ Hlawatsch
are with the Institute of Telecommunications, TU Wien, 1040 Vienna, Austria
(email: \{fmeyer,$\ist$fhlawats\}@nt.tuwien.ac.at).
O.\ Hlinka is with robart GmbH, 4020 Linz, Austria (e-mail: ondrej.hlinka@robart.cc).
H.\ Wymeersch is with the Department of Signals and Systems, Chalmers University of Technology, Gothenburg 41296, Sweden (email: henk.wymeersch@ieee.org).
E.\ Riegler is with the Department of Information Technology and Electric\-al Engineering, ETH Zurich, 8092 Zurich, Switzerland (email: eriegler@nari.ee.ethz.ch).
This work was supported by the FWF under Grants S10603-N13 and P27370-N30,
by the WWTF under Grant ICT10-066 (NOWIRE), and by the European Commission under
ERC Grant No. 258418 (COOPNET), the Newcom\# Network of Excellence in Wireless Communications, 
and the National Sustainability Program (Grant LO1401).
This work was partly presented at the 46th Asilomar Conference on Signals, Systems and Computers, Pacific Grove, CA, Nov.\ 2012.}}

\maketitle

\begin{abstract}
We propose a Bayesian method for distributed sequential localization of mobile networks composed of both cooperative agents and noncooperative objects. 
Our method provides a consistent combination of cooperative self-localization (CS) and distributed tracking (DT). Multiple mobile agents and objects are localized and tracked 
using measurements between agents and objects and between agents. For a distributed operation and low complexity, 
we combine particle-based belief propagation with a consensus or gossip scheme. High localization accuracy is achieved through a probabilistic 
information transfer between the CS and DT parts of the underlying factor graph. Simulation results demonstrate significant improvements in both 
agent self-localization and object localization performance compared to separate CS and DT, and very good scaling properties with respect to the numbers of agents and objects.
\end{abstract}

\begin{IEEEkeywords}
Agent network, 
belief propagation, 
consensus,
cooperative localization, 
distributed estimation, 
distributed tracking, 
factor graph, 
gossip,
message passing, 
sensor network.
\end{IEEEkeywords}

\section{Introduction}\label{sec:intro}

\vspace{.5mm}

\subsection{Background and State of the Art}
\label{sec:intro_background}


Cooperative self-localization (CS) \cite{patwari,wymeersch} and distributed tracking (DT) \cite{liu07} are key signal processing tasks in decentralized agent networks. 
Applications include surveillance \cite{aghajan2009multi}, environmental and agricultural monitoring \cite{corke10}, robotics \cite{bullo09}, and pollution 
source localization \cite{nayak2010wireless}. In CS, each agent measures quantities related to the location of neighboring agents relative to its own location.
By cooperating with other agents, it is able to estimate its own location. In DT, the measurements performed by the agents are related to the 
locations (or, more generally, states) of noncooperative objects to be tracked. 
At each agent, estimates of the 
\pagebreak 
object states are cooperatively calculated from all agent measurements. CS and  DT are related since, ideally, an agent needs to know its 
own location to be able to contribute to DT. This relation motivates the combined CS-DT method proposed in this paper, 
which achieves improved performance through a probabilistic information transfer between CS and DT.

For CS of static agents (hereafter termed ``static CS''), the nonparametric belief propagation (BP) algorithm has been proposed in \cite{ihler}. 
BP schemes are well suited to CS because their complexity scales only linearly with the number of agents and, under mild assumptions, 
a distributed implementation is easily obtained.
In \cite{wymeersch}, a distributed BP message passing algorithm for CS of mobile agents (hereafter termed ``dynamic CS'') is proposed. 
A message passing algorithm based on the mean field approximation is presented for static CS in \cite{pedersen}. 
In \cite{lien} and \cite{li15}, nonparametric BP is extended to dynamic CS and combined with a parametric message representation.
In \cite{savic12reduc}, a particle-based BP method using a Gaussian belief approximation is proposed. 
The low-complexity method for dynamic CS presented in \cite{sathyan13} is based on the Bayesian filter and a linearized measurement equation. Another low-complexity CS
method with low communication requirements is sigma point BP \cite{meyer14sigma}. In \cite{vandevelde15}, a censoring scheme for 
sigma point BP is proposed to further reduce communications.

For DT, various distributed recursive estimation methods are available, e.g., \cite{mutambara98, vercauteren05, hlinkaMag13,farahmand11,hlinka14adaptation}.
Distributed particle filters \cite{hlinkaMag13} are especially attractive since they are suited to nonlinear, non-Gaussian systems. In particular, 
in the distributed particle filters proposed in \cite{farahmand11} and \cite{hlinka14adaptation}, consensus algorithms are used to com\-pute
global particle weights reflecting the measurements of all agents. 
For DT of an unknown, possibly time-varying number of objects in the presence of object-to-measurement association uncertainty, methods 
based on random finite sets are proposed in \cite{uney13,battistelli13,fantacci15}.
 
In the framework of \emph{simultaneous localization and tracking} (SLAT), static agents track a noncooperative object and localize themselves, 
using measurements of the distances between each agent and the object \cite{taylor}. In contrast to dynamic CS, measurements between agents are only used for initialization.
A centralized particle-based SLAT method using BP is proposed in \cite{savic15}. 
Distributed SLAT methods include a technique using a Bayesian filter and communication via a junction tree \cite{funiak}, 
iterative maximum likelihood methods \cite{kantas12}, variational filtering \cite{teng2012distr}, and particle-based BP \cite{uney14}.

\vspace{-1mm}

\subsection{Contributions and Paper Organization}

We propose a method for distributed localization and tracking of cooperative agents and noncooperative objects in wireless networks, using 
measurements between agents and objects and between agents. This method, for the first time, provides a consistent combination of CS and DT 
in decentralized agent networks where the agents and objects may be mobile. To the best of our knowledge, it
is the first method for simultaneous CS and DT in a dynamic setting. It is different from SLAT methods in that the agents may be mobile and 
measurements between the agents are used also during runtime. A key feature of our method is a 
probabilistic information transfer between the CS and DT stages, which allows uncertainties 
in one stage to be taken into account by the other stage and thereby can improve the performance of both stages. 

Contrary to the multitarget tracking literature \cite{barShalom11, mahler2007statistical}, we assume that the number of objects is known and the objects 
can be identified by the agents. Even with this assumption, the fact that the agents may be mobile and their states  
are unknown causes the object localization problem to be more challenging than in the setting of static agents with known states. 
This is because the posterior distributions of the object and agent states are coupled through recurrent pairwise measurements, and thus all these states 
should be estimated \emph{jointly} and \emph{sequentially}. This joint, sequential estimation is performed quite naturally through our factor graph formulation 
of the entire estimation problem and the use of BP message passing. In addition, BP message passing facilitates a distributed implementation and exhibits 
very good scalability in the numbers of agents and objects. We also present a new particle-based implementation of BP message passing that is less complex 
than conventional nonparametric BP \cite{ihler, lien}.

Our method is an extension of BP-based dynamic CS \cite{wymeersch,lien,li15} to include noncooperative objects. 
This extension is nontrivial because, contrary to pure CS, the communication and measurement topologies (graphs) do not match. Indeed, 
because the objects do not communicate, certain messages needed to calculate the object beliefs are not available at the agents. The proposed method 
employs a consensus scheme \cite{farahmand11} for a distributed calculation of these messages. The resulting combination of BP and consensus 
may also be useful in other distributed inference problems involving noncooperative objects.

This paper is organized as follows. The system model is described in Section \ref{sec:sceass}. A BP message passing scheme for joint CS and DT 
is developed in Section \ref{sec:slat}, and a particle-based implementation of this scheme in Section \ref{sec:NBP_CoSLAT}. 
A distributed localization-and-tracking algorithm that combines particle-based BP and consensus is presented in Section \ref{sec:slat_algor}. 
Variations and implementation aspects of the proposed algorithm are discussed in Section \ref{sec:varImp}. The algorithm's communication requirements and delay are analyzed in Section \ref{sec:CoSLAT_comm}. Finally, simulation results are presented in Section \ref{sec:simres}.

This paper extends our previous work 
in \cite{meyer12,meyer2013coop} in the following respects: 
we consider multiple objects, 
introduce a new low-complexity message multiplication scheme,
provide an analysis of communication requirements and delay,
present an alternative particle-based processing that allows for uninformative prior distributions, 
introduce a modification of the consensus scheme that yields fast convergence,
present a numerical comparison with particle filtering
and a numerical analysis of scaling properties,
and demonstrate performance gains over separate CS and DT in a static scenario (in addition to two dynamic scenarios).
We note that an extension of the proposed method that includes a distributed information-seeking controller optimizing the behavior (e.g., movement) of the agents is presented in \cite{meyer2015infoseeking}.

\section{System Model and Problem Statement}\label{sec:sceass}

\vspace{1mm}

\subsection{System Model}
\label{sec:sceass:systmod}

\begin{figure}[t!]
\vspace{1mm}
\centering
\psfrag{S1}[l][l][.76]{\raisebox{2mm}{\hspace{0mm}cooperative agent}}
\psfrag{S2}[l][l][.76]{\raisebox{-3mm}{\hspace{0mm}noncooperative object}}
\psfrag{S3}[l][l][.76]{\raisebox{1mm}{\hspace{0mm}communication link}}
\psfrag{S4}[l][l][.76]{\raisebox{.5mm}{\hspace{0mm}measurement}}
\psfrag{A5}[l][l][.76]{\raisebox{.5mm}{\hspace{0mm}}}
\psfrag{A3}[l][l][.70]{\raisebox{6mm}{\hspace{-3.7mm}$\cl{M}^{\OO}_{l,n}$}}
\psfrag{A2}[l][l][.70]{\raisebox{.5mm}{\hspace{-1mm}$\cl{M}^{\AA}_{l,n}$}}
\psfrag{A4}[l][l][.70]{\raisebox{.5mm}{\hspace{0mm}$\cl{M}_{l,n}$}}
\psfrag{A1}[l][l][.70]{\raisebox{12.3mm}{\hspace{-11mm}agent $l$}}
\psfrag{A10}[l][l][.70]{\raisebox{6.1mm}{\hspace{-7.5mm}agent $l'$}}
\psfrag{A6}[l][l][.70]{\raisebox{-6.3mm}{\hspace{-3.3mm}object}}
\psfrag{A7}[l][l][.70]{\raisebox{-4.8mm}{\hspace{0mm}$m$}}
\psfrag{A9}[l][l][.70]{\raisebox{1mm}{\hspace{0mm}$\cl{C}_{l'\!,n}$}}
\psfrag{A8}[l][l][.70]{\raisebox{1mm}{\hspace{-1.2mm}$\AA_{m,n}$}}
\hspace*{-18mm}\includegraphics[scale=.67]{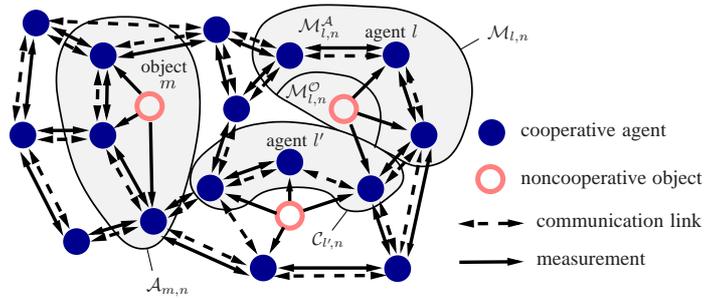}
\vspace{-2.5mm}
\caption{Network with cooperative agents and noncooperative objects. Also shown are the sets $\cl{M}_{l,n}$, $\cl{M}^{\AA}_{l,n}$, and $\cl{M}^{\OO}_{l,n}$ 
for a specific agent $l$, the set $\cl{C}_{l'\!,n}$ for a specific agent $l'\!$, and the set $\AA_{m,n}$ for a specific object $m$.}
\label{fig:an}
\vspace{-2mm}
\end{figure}

We consider a decentralized network of cooperative agents and noncooperative objects as shown
in Fig.\ \ref{fig:an}. We denote by $\AA \rmv\subseteq\rmv \mathbb{N}$ the set of agents,
by $\OO \rmv\subseteq\rmv \mathbb{N}$ the set of objects, and by $\EE \triangleq \AA \cup \OO$ the set of all entities (agents and objects). We 
use the indices $k \rmv\in\! \EE$, $l \rmv\in\! \AA$, and $m \rmv\in\! \OO$  to denote a generic entity, an agent, and an object, respectively.
The numbers of agents and objects are assumed known. The objects are noncooperative in that they do not communicate, do not perform computations, 
and do not actively perform any measurements. The \emph{state} of entity $k \rmv\in\! \EE$ at time $n \rmv\in\rmv \{0,1,\ldots\}$, denoted $\bd{x}_{k,n}$, 
consists of the current location and, possibly, motion parameters such as velocity \cite{rong}. The states evolve according 
\vspace{-1.5mm}
to 
\begin{equation}
\label{eq_statrans}
\bd{x}_{k,n} =\ist g(\bd{x}_{k,n-1},\bd{u}_{k,n}) \,, \quad k \rmv\in\! \EE \,,
\end{equation}
where $\bd{u}_{k,n}$ denotes driving noise with probability density function (pdf) $f(\bd{u}_{k,n})$. The statistical relation between 
$\bd{x}_{k,n-1}$ and $\bd{x}_{k,n}$ defined by \eqref{eq_statrans} can also be described by the state-transition pdf $f(\bd{x}_{k,n}|\bd{x}_{k,n-1})$. 

The communication and measurement topologies are described by sets $\cl{C}_{l,n}$ and $\cl{M}_{l,n}$ as follows. 
Agent $l \rmv\in\rmv \AA$ is able to communicate with agent $l'$ if $l'\rmv\in \cl{C}_{l,n}\rmv\subseteq\rmv \AA \rmv\setminus\rmv \{l\}$. 
Communication is symmetric, i.e., $l'\!\in\rmv \cl{C}_{l,n}$ implies $l \rmv\in\rmv \cl{C}_{l'\rmv,n}$.
Furthermore, agent $l \in \AA$ acquires a measurement $\bd{y}_{l,k;n}$ relative to agent or object $k \rmv\in\rmv \EE$ 
if $k \!\in\rmv \cl{M}_{l,n} \rmv\subseteq\rmv \EE \setminus\rmv \{l\}$. Note that $\cl{C}_{l,n}$ consists of all agents that can communicate 
with agent $l \in \AA$, and $\cl{M}_{l,n}$ consists of all entities measured by agent $l \in \AA$. Thus, there are actually two networks:
one is defined by the \emph{communication graph}, i.e., by $\cl{C}_{l,n}$ for $l \in \AA$, and involves only agents; the other 
is defined by the \emph{measurement graph}, i.e., by $\cl{M}_{l,n}$ for $l \in \AA$, and involves agents and objects.
The communication graph is assumed to be connected. We also define $\cl{M}^{\AA}_{l,n} \!\triangleq\rmv \cl{M}_{l,n} \cap \AA$ and 
$\cl{M}^{\OO}_{l,n} \!\triangleq\rmv \cl{M}_{l,n} \cap \OO$, i.e., the subsets of $\cl{M}_{l,n}$ containing only agents and only objects, respectively, 
and $\AA_{m,n} \!\triangleq\rmv \{ l\rmv\in\rmv \AA \ist |\ist  m \rmv\in\rmv \cl{M}_{l,n}^{\OO}\}$, i.e., the set of agents that acquire measurements of object $m$. 
Note that $m \rmv\in\rmv \cl{M}_{l,n}^{\OO}$ if and only if $l \rmv\in\rmv \AA_{m,n}$.
We assume that $\cl{M}^{\AA}_{l,n} \!\subseteq\rmv \cl{C}_{l,n}$, i.e., if agent $l$ acquires a measurement relative to agent $l'$, 
it is able to communicate with agent $l'$. The sets $\cl{C}_{l,n}$ etc.\ may be time-dependent. An example of communication and 
measurement topologies is given in Fig.\ \ref{fig:an}. 

We consider ``pairwise'' measurements $\bd{y}_{l,k;n}$ that depend on the states $\bd{x}_{l,n}$ and $\bd{x}_{k,n}$ according to
\begin{equation}
\label{eq:meas_mod}
\bd{y}_{l,k;n} =\ist h(\bd{x}_{l,n}, \bd{x}_{k,n}, \bd{v}_{l,k;n}) \,, \quad l \rmv\in\rmv \AA \ist, \; k \rmv\in\rmv \cl{M}_{l,n} \,.
\end{equation}
Here, $\bd{v}_{l,k;n}$ is measurement noise with pdf $f(\bd{v}_{l,k;n})$. An example is the scalar ``noisy distance'' measurement
\begin{equation}
\label{eq:mess}
y_{l,k;n} = \|\tilde{\bd{x}}_{l,n} \!-\rmv \tilde{\bd{x}}_{k,n} \| + v_{l,k;n} \,,
\end{equation}
where $\tilde{\bd{x}}_{k,n}$ represents the location of entity $k$ (this is part of the state $\bd{x}_{k,n}$). The statistical dependence of $\bd{y}_{l,k;n}$ on
$\bd{x}_{l,n}$ and $\bd{x}_{k,n}$ is described by the local likelihood function $f(\bd{y}_{l,k;n}|\bd{x}_{l,n},\bd{x}_{k,n})$. 
We denote by $\bd{x}_{n} \rmv\triangleq {\big( \bd{x}_{k,n} \big)}_{k \in \EE}\rmv$ and $\bd{y}_{n} \rmv\triangleq$\linebreak 
${\big( \bd{y}_{l,k;n} \big)}_{l \in \AA, \, k \in \cl{M}_{l,n}}\!$ the vectors of, respectively, all states and measurements 
at time $n$. Furthermore, we define $\bd{x}_{1:n} \rmv\triangleq \big( \bd{x}^{\text{T}}_{1} \cdots\ist \bd{x}^{\text{T}}_{n}\big)^{\text{T}}$ 
and $\bd{y}_{1:n} \rmv\triangleq \big( \bd{y}^{\text{T}}_{1} \cdots\ist \bd{y}^{\text{T}}_{n}\big)^{\text{T}}\rmv$.

\vspace{-1.5mm}

\subsection{Assumptions}
\label{sec:sceass:assum}

We will make the following commonly used assumptions, which are reasonable in many practical scenarios \cite{wymeersch}.

(A1) All agent and object states are independent \emph{a priori} at time $n \!=\! 0$, i.e., $f(\bd{x}_{0}) = \prod_{k \in \EE} f(\bd{x}_{k,0})$.

(A2) All agents and objects move according to a memoryless walk, i.e., $f(\bd{x}_{1:n}) = f(\bd{x}_{0}) \prod_{n' = 1}^{n} f(\bd{x}_{n'}|\bd{x}_{n'-1})$.

(A3) The state transitions of the various agents and objects are independent, i.e., $f(\bd{x}_{n}|\bd{x}_{n-1}) = \prod_{k \in\ist \EE} f(\bd{x}_{k,n}|\bd{x}_{k,n-1})$.

(A4) The current measurements $\bd{y}_{n}$ are conditionally independent, given the current states $\bd{x}_{n}$, 
of all the other states and of all past and future measurements, i.e., $f(\bd{y}_{n}|\bd{x}_{0:\infty}, \bd{y}_{1:n-1},$\linebreak 
$\bd{y}_{n+1:\infty}) = f(\bd{y}_{n}|\bd{x}_{n})$.

(A5) The current states $\bd{x}_{n}$ are conditionally independent of all past measurements, $\bd{y}_{1:n-1}$, 
given the previous states $\bd{x}_{n-1}$, i.e., $f(\bd{x}_{n}|\bd{x}_{n-1}, \bd{y}_{1:n-1}) = f(\bd{x}_{n}|\bd{x}_{n-1})$.

(A6) The measurements $\bd{y}_{l,k;n}$ and $\bd{y}_{l'\!,k';n}$ are conditionally independent given $\bd{x}_{n}$ unless $(l,k) \!=\! (l'\!,k')$, and each
measurement $\bd{y}_{l,k;n}$ depends only on the states $\bd{x}_{l,n}$ and $\bd{x}_{k,n}$.
Together with (A4), this leads to the following factorization of the ``total'' likelihood function: 
$f(\bd{y}_{1:n}|\bd{x}_{1:n}) = \prod_{n'=1}^{n}\prod_{l \in \AA} \prod_{k \in \cl{M}_{l,n}} \! f(\bd{y}_{l,k;n'}|\bd{x}_{l,n'},\bd{x}_{k,n'})$.

We also assume that the objects can be identified by the agents, i.e., object-to-measurement associations are known. 
(We note that BP-based methods for multitarget tracking in the presence of object-to-measurement association uncertainty were recently 
proposed in \cite{williams14} and \cite{meyer15scalable}; however, these methods are not distributed.)
Furthermore, we assume that each agent $l \rmv\in\rmv \AA$ knows the functional forms of its own state-transition pdf and initial state pdf
as well 
\pagebreak 
as of those of all objects, i.e., $f(\bd{x}_{k,n}|\bd{x}_{k,n-1})$ and $f(\bd{x}_{k,0})$ for $k \in \{l\} \cup \OO$. 
Finally, all prior location and motion information is available in one global reference frame, 
and the internal clocks of all agents are synchronous (see \cite{wu11,etzlinger14,meyerAsilomar13} for distributed clock synchronization algorithms).

\vspace{-1mm}

\subsection{Problem Statement}
\label{sec:sceass:prob}

The task we consider is as follows: Each agent $l \rmv\in\rmv \AA$ estimates its own state $\bd{x}_{l,n}$ and all object states
$\bd{x}_{m,n}$, $m \rmv\in\rmv \OO\rmv$ from the \emph{entire measurement vector}
$\bd{y}_{1:n} \rmv= {\big( \bd{y}_{l'\!,k;n'} \big)}_{l' \in \AA, \, k \in \cl{M}_{l'\!,n'}, n' \in \{1,\ldots,n\}}$, 
\vspace{.5mm}
i.e., from the pairwise measurements between the agents and between the agents and objects up to time $n$.
This is to be achieved using only communication with the ``neighbor agents'' as defined by $\cl{C}_{l,n}$, and without
transmitting measurements between agents.

In this formulation of the estimation task, compared to pure CS of cooperative agents (e.g., \cite{wymeersch}) or
pure DT of noncooperative objects (e.g., \cite{hlinka14adaptation}), the measurement set is extended in that it includes also the respective other 
measurements---i.e., the measurements between agents and objects for agent state estimation and those between agents for object state estimation. 
This explains why the proposed algorithm is able to outperform separate CS and DT. In fact, by using 
all the present and past measurements available throughout the entire network, the inherent coupling between the CS and DT tasks 
can be exploited for improved performance.

\section{BP Message Passing Scheme} 
\label{sec:slat}

\vspace{.5mm}

In this section, we describe a BP message passing scheme for the joint CS-DT problem. 
A particle-based implementation of this scheme will be presented in Section \ref{sec:NBP_CoSLAT},
and the final distributed algorithm will be developed in Section \ref{sec:slat_algor}.

For estimating the agent or object state $\bd{x}_{k,n}$, $k \rmv\in\rmv \EE$ from $\bd{y}_{1:n}$, a popular Bayesian estimator is the 
minimum mean-square error (MMSE) estimator  given by \cite{kay}
\be
\hat{\bd{x}}^\text{MMSE}_{k,n} \,\triangleq \int \rmv \bd{x}_{k,n} \ist f(\bd{x}_{k,n}|\bd{y}_{1:n}) \ist d\bd{x}_{k,n} \,.
\label{eq:mmse_slat}
\ee
This estimator involves the ``marginal'' posterior pdf $f(\bd{x}_{k,n}|\bd{y}_{1:n})$, $k \rmv\in\rmv \EE$, which can be obtained by marginalizing 
the ``joint'' posterior pdf $f(\bd{x}_{1:n}|\bd{y}_{1:n})$. However, direct marginalization of $f(\bd{x}_{1:n}|\bd{y}_{1:n})$ is infeasible
because it relies on nonlocal information and involves integration in spaces whose dimension grows with time and network size. This problem can be addressed by
using a particle-based, distributed BP scheme that takes advantage of the temporal and spatial independence structure of $f(\bd{x}_{1:n}|\bd{y}_{1:n})$
and avoids explicit integration. The independence structure corresponds to the following factorization of $f(\bd{x}_{1:n}|\bd{y}_{1:n})$,
which is obtained by using Bayes' rule and assumptions (A1)--(A6):
\begin{align}
\hspace{-2mm}f(\bd{x}_{1:n}|\bd{y}_{1:n})  \ist\propto\, \ist& \Bigg( \rmv\prod_{k \in \EE} \rmv f(\bd{x}_{k,0}) \rmv\Bigg) \nonumber \\
  &\hspace{.1mm}\times \rmv\rmv\prod_{n' = 1}^n \!\Bigg( \rmv\prod_{k_1 \in \EE} \rmv f(\bd{x}_{k_1,n'}|\bd{x}_{k_1,n'-1}) \rmv\Bigg) \nonumber \\[1mm]
  & \hspace{.1mm}\times \prod_{l \in \AA} \ist \prod_{k_2 \in \cl{M}_{l,n'}} \!\! \rmv\rmv 
  f(\bd{y}_{l,k_2;n'}|\bd{x}_{l,n'},\bd{x}_{k_2,n'}) \,. \label{eq:post_pdf}
  \\[-6.5mm]
\nonumber
\end{align}
Here, $\propto$ denotes equality up to a constant normalization factor.

\begin{figure}
\vspace{2mm}
\centering
\psfrag{S17}[l][l][.8]{\raisebox{-2.3mm}{\hspace{-0.7mm}$\bd{x}_{1}$}}
\psfrag{S21}[l][l][.8]{\raisebox{-2.3mm}{\hspace{-0.7mm}$\bd{x}_{2}$}}
\psfrag{S24}[l][l][.8]{\raisebox{-2mm}{\hspace{-1.4mm}$\bd{x}_{L}$}}
\psfrag{S3}[l][l][.8]{\raisebox{-2.3mm}{\hspace{-0.7mm}$\bd{x}_{1}$}}
\psfrag{S7}[l][l][.8]{\raisebox{-2.3mm}{\hspace{-0.7mm}$\bd{x}_{2}$}}
\psfrag{S12}[l][l][.8]{\raisebox{-2mm}{\hspace{-1.7mm}$\bd{x}_{L}$}}
\psfrag{S2}[l][l][.8]{\raisebox{-3mm}{\hspace{-0.7mm}$f_1$}}
\psfrag{S18}[l][l][.8]{\raisebox{-3mm}{\hspace{-0.7mm}$f_1$}}
\psfrag{S6}[l][l][.8]{\raisebox{-3mm}{\hspace{-0.7mm}$f_2$}}
\psfrag{S20}[l][l][.8]{\raisebox{-3mm}{\hspace{-0.7mm}$f_2$}}
\psfrag{S13}[l][l][.8]{\raisebox{1mm}{\hspace{-1.5mm}$f_L$}}
\psfrag{S23}[l][l][.8]{\raisebox{-2mm}{\hspace{-1.5mm}$f_L$}}
\psfrag{S5}[l][l][.8]{\raisebox{1.5mm}{\hspace{0.3mm}$\tilde{f}_1$}}
\psfrag{S8}[l][l][.8]{\raisebox{1.5mm}{\hspace{0.3mm}$\tilde{f}_2$}}
\psfrag{S11}[l][l][.8]{\raisebox{1.5mm}{\hspace{-0.3mm}$\tilde{f}_L$}}
\psfrag{S16}[l][l][.9]{\raisebox{3mm}{\hspace{-3mm}$n \!-\! 1$}}
\psfrag{S15}[l][l][.9]{\raisebox{3mm}{\hspace{-2mm}$n$}}	
\psfrag{S4}[l][l][.8]{\raisebox{-4mm}{\hspace{-2.5mm}$f_{1,2}$}}
\psfrag{S31}[l][l][.8]{\raisebox{-3mm}{\hspace{-1mm}$f_{2,1}$}}
\psfrag{S33}[l][l][.8]{\raisebox{-3mm}{\hspace{-1mm}$f_{2,1}$}}
\psfrag{S32}[l][l][.8]{\raisebox{-4mm}{\hspace{-1.8mm}$f_{1,l}$}}
\psfrag{S34}[l][l][.8]{\raisebox{-4mm}{\hspace{-1.2mm}$f_{1,l}$}}
\psfrag{S35}[l][l][.8]{\raisebox{-1mm}{\hspace{-2mm}$f_{L,l}$}}
\psfrag{S36}[l][l][.8]{\raisebox{-1mm}{\hspace{-2mm}$f_{L,l}$}}
\psfrag{S19}[l][l][.8]{\raisebox{-3mm}{\hspace{-2.4mm}$f_{1,2}$}}
\psfrag{S9}[l][l][.8]{\raisebox{0mm}{\hspace{-1mm}$f_{2,l}$}}
\psfrag{S22}[l][l][.8]{\raisebox{0mm}{\hspace{-1mm}$f_{2,l}$}}
\psfrag{S28}[l][l][.65]{\raisebox{2.3mm}{\hspace{-1.7mm}$b_1^{(P)}$}}
\psfrag{S29}[l][l][.65]{\raisebox{1.8mm}{\hspace{-3.4mm}$b_2^{(p-1)}$}}
\psfrag{S30}[l][l][.65]{\raisebox{1.8mm}{\hspace{-3.5mm}$b_{l}^{(p-1)}$}}
\psfrag{S26}[l][l][.7]{\raisebox{2.3mm}{\hspace{-3.9mm}$\phi_{l\rightarrow 1}^{(p)}$}}
\psfrag{S25}[l][l][.65]{\raisebox{2mm}{\hspace{-.5mm}$\phi_{2\rightarrow 1}^{(p)}$}}
\psfrag{S27}[l][l][.7]{\raisebox{-1.5mm}{\hspace{-2.3mm}$\phi_{\rightarrow n}$}}
\psfrag{S45}[l][l][.8]{\raisebox{-2.2mm}{\hspace{-1.2mm}$\bd{x}_{m}$}}
\psfrag{S46}[l][l][.8]{\raisebox{-2.5mm}{\hspace{-1.2mm}$\bd{x}_{m}$}}
\psfrag{S43}[l][l][.8]{\raisebox{-2mm}{\hspace{-0.7mm}$f_{m}$}}
\psfrag{S44}[l][l][.8]{\raisebox{-1mm}{\hspace{-0.7mm}$f_{m}$}}
\psfrag{S40}[l][l][.8]{\raisebox{0mm}{\hspace{-3mm}$f_{2,m}$}}
\psfrag{S41}[l][l][.8]{\raisebox{0mm}{\hspace{-3.1mm}$f_{1,m}$}}
\psfrag{S42}[l][l][.8]{\raisebox{0mm}{\hspace{-4mm}$f_{L,m}$}}
\psfrag{S49}[l][l][.8]{\raisebox{0mm}{\hspace{-4mm}$f_{L,m}$}}
\psfrag{S47}[l][l][.7]{\raisebox{0mm}{\hspace{-1.7mm}$\phi_{\rightarrow n}$}}
\psfrag{S48}[l][l][.7]{\raisebox{0mm}{\hspace{-3.4mm}}}
\psfrag{S47}[l][l][.7]{\raisebox{1mm}{\hspace{-0.6mm}$\phi_{1 \rightarrow m}^{(p)}$}}
\psfrag{S50}[l][l][.7]{\raisebox{1.5mm}{\hspace{-9mm}}}
\psfrag{S51}[l][l][.7]{\raisebox{4.5mm}{\hspace{-1.5mm}$\psi_{1 \rightarrow m}^{(p-1)}$}}
\psfrag{S52}[l][l][.7]{\raisebox{3mm}{\hspace{-4mm}$\phi_{l \rightarrow m}^{(p)}$}}
\psfrag{S53}[l][l][.7]{\raisebox{1.5mm}{\hspace{-1mm}$\psi_{m \rightarrow 1}^{(p-1)}$}}
\psfrag{S54}[l][l][.7]{\raisebox{3mm}{\hspace{-2.5mm}$\phi_{m \rightarrow 1}^{(p)}$}}
\psfrag{S55}[l][l][.7]{\raisebox{2mm}{\hspace{-3mm}$\phi_{\rightarrow n}$}}
\psfrag{S56}[l][l][.7]{\raisebox{-5mm}{\hspace{-2.5mm}$b_{m}^{(P)}$}}
\hspace{-1mm}\includegraphics[scale=0.31]{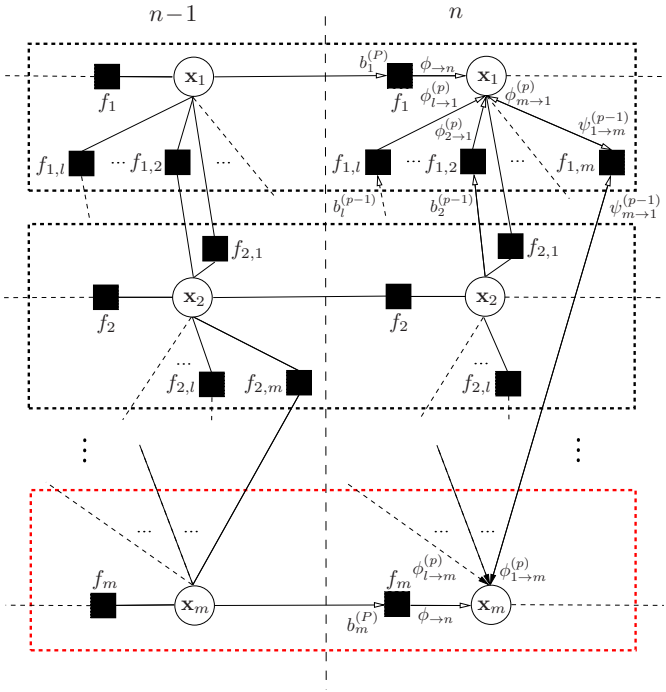}
\renewcommand{\baselinestretch}{1.2}\small\normalsize
\vspace{-3mm}
\caption{Factor graph showing the states of agents $l \!=\! 1$ and $l \!=\! 2$ and of a single object $m$ at time instants $n\!-\!1$ and $n$, 
assuming $2 \rmv\in\rmv \cl{C}_{1,n}$ and $m \rmv\in \cl{M}^{\OO}_{1,n} \rmv\cap \cl{M}^{\OO}_{2,n-1}$. 
Variable and factor nodes are depicted as circles and squares, respectively. Time indices are omitted for simplicity.  
The short notation $f_{k} \triangleq f(\bd{x}_{k,n'}|\bd{x}_{k,n'-1})$, $f_{l,k} \triangleq f(\bd{y}_{l,k;n'} | \bd{x}_{l,n'},\bd{x}_{k,n'})$, 
$b_k^{(p)} \!\triangleq b^{(p)}(\bd{x}_{k,n'})$,
$\psi_{k \rightarrow k'}^{(p)} \rmv\triangleq\rmv \psi^{(p)}_{k \rightarrow k'}(\bd{x}_{k,n'})$, etc.\ (for $n' \!\in\rmv \{1, \dots, n\}$) 
is used. The upper two (black) dotted boxes correspond to the CS part (for agents $l \!=\! 1,2$); the bottom (red) dotted box corresponds to the DT part. 
Edges between black dotted boxes imply communication between agents. Only messages and beliefs involved in the computation of 
$b^{(p)}(\bd{x}_{1,n})$  and $b^{(p)}(\bd{x}_{m,n})$ are shown. Edges with non-filled arrowheads depict particle-based messages and beliefs, 
while edges with filled arrowheads depict messages involved in the consensus scheme.}
\label{fig:coslat_fg}
\vspace{.5mm}
\end{figure}

\vspace{-1mm}

\subsection{The Message Passing Scheme}
\label{sec:messpass:scheme}

The proposed BP message passing scheme uses the sum-product algorithm \cite{kschischang} to produce
approximate marginal posterior pdfs (``beliefs'') $b(\mathbf{x}_{k,n}) \approx f(\bd{x}_{k,n}|\bd{y}_{1:n})$, $k \rmv\in\rmv \EE$.
This is based on the factor graph \cite{loeliger} corresponding to the factorization of $f(\bd{x}_{1:n}|\bd{y}_{1:n})$ in \eqref{eq:post_pdf}, which is shown
in Fig.\ \ref{fig:coslat_fg}. The factor graph contains variable nodes, factor nodes, and edges connecting certain variable nodes with certain factor nodes. 
Because the factor graph is loopy (i.e., some edges form loops), BP schemes provide only an approximate marginalization. However, the resulting beliefs 
have been observed to be quite accurate in many applications \cite{kschischang, wainwright, wymeersch}.
In loopy factor graphs, the BP scheme becomes iterative, and there exist different orders in which messages can be computed 
\cite{kschischang, wainwright, wymeersch}. 

Here, we choose an order that enables real-time processing and facilitates a distributed implementation.
More specifically, the belief of agent node $l \rmv\in\rmv \AA$ and object node $m \rmv\in\rmv \OO$ at time $n$ and message passing iteration 
$p \rmv\in\rmv \{1,\ldots,P\}$ is given
\vspace*{.5mm} 
by
\begin{align}
\hspace*{-2mm}b^{(p)}(\bd{x}_{l,n}) &\,\propto\, \phi_{\rightarrow n}(\bd{x}_{l,n}) \!\!\!\prod_{\,k \in \cl{M}_{l,n}} \!\!\rmv 
  \phi^{(p)}_{k \rightarrow l}(\bd{x}_{l,n}) \,, \hspace{2mm} l \!\in\! \AA  \label{eq:marg_CoSLAT_agent} 
 \\[.5mm]
\hspace*{-2mm}b^{(p)}(\bd{x}_{m,n}) &\,\propto\, \phi_{\rightarrow n}(\bd{x}_{m,n}) \!\!\prod_{\,l \in \AA_{m,n}} \!\!\rmv 
  \phi^{(p)}_{l \rightarrow m}(\bd{x}_{m,n}) \,, \hspace{2mm}  m \!\in\! \OO \ist, \label{eq:marg_CoSLAT_T}\\[-6mm]
  \nonumber
\end{align}
respectively, with the ``prediction message''
\begin{equation}
\label{eq:pred_mess}
\hspace*{-.1mm}\phi_{\rightarrow n}(\bd{x}_{k,n}) =\! \int \rmv\rmv f(\bd{x}_{k,n}|\bd{x}_{k,n-1}) \ist b^{(P)}(\bd{x}_{k,n-1}) \ist d\bd{x}_{k,n-1} \ist, 
  \;\, k \!\in\! \EE 
\vspace{-1mm} 
\end{equation}
and the ``measurement messages''
\begin{align}
\phi_{k \rightarrow l}^{(p)}(\bd{x}_{l,n}) &\,= \begin{cases} 
\begin{array}[t]{lr} \hspace{-2mm}\int \rmv f(\bd{y}_{l,k;n} | \bd{x}_{l,n}\ist,\bd{x}_{k,n}) \, b^{(p-1)}(\bd{x}_{k,n}) \, d\bd{x}_{k,n} \,, \\[1mm]
  \hspace{36mm} k \!\in\! \cl{M}^{\AA}_{l,n} \ist,\; l \!\in\! \AA \end{array}\\[8mm]
\begin{array}[t]{lr}\hspace{-2mm}\int \rmv f(\bd{y}_{l,k;n} | \bd{x}_{l,n}\ist,\bd{x}_{k,n}) \, \psi^{(p-1)}_{k \rightarrow l}(\bd{x}_{k,n}) \, 
  d\bd{x}_{k,n} \,,\\[1mm]  
  \hspace{36mm} k \!\in\! \cl{M}^{\OO}_{l,n} \ist,\; l \!\in\! \AA 
\end{array}
\end{cases} \nonumber\\[0mm]
\label{eq:meas_mess_1}\\[-10mm]
\nonumber
\end{align}
and
\begin{align}
\hspace{-2.1mm}\phi_{l \rightarrow m}^{(p)}(\bd{x}_{m,n}) &\,= \int \rmv f(\bd{y}_{l,m;n} | \bd{x}_{l,n}\ist,\bd{x}_{m,n}) \,  \psi^{(p-1)}_{l \rightarrow m}(\bd{x}_{l,n}) \, d\bd{x}_{l,n}\,, \nonumber\\[-1mm]
  &\hspace{34mm} l \!\in\! \AA_{m,n} \ist,\; m \!\in\! \OO \ist. \label{eq:meas_mess_3}\\[-6mm]
\nonumber
\end{align}
Here, $\psi^{(p-1)}_{m \rightarrow l}(\bd{x}_{m,n})$ and $\psi^{(p-1)}_{l \rightarrow m}(\bd{x}_{l,n})$ (constituting the ``extrinsic information'') are given 
\vspace{.5mm}
by 
\begin{align}
\hspace{-2.1mm}\psi^{(p-1)}_{m \rightarrow l}(\bd{x}_{m,n}) &\eq
     \phi_{\rightarrow n}(\bd{x}_{m,n}) \!\!\prod\limits_{l' \in \AA_{m,n} \rmv\backslash \{l\}} \!\! \phi^{(p-1)}_{l' \rightarrow m}(\bd{x}_{m,n}) 
       \label{eq:extr_infor_2}\\[1mm]
\hspace{-2.1mm}\psi^{(p-1)}_{l \rightarrow m}(\bd{x}_{l,n}) &\eq
     \phi_{\rightarrow n}(\bd{x}_{l,n}) \!\!\prod\limits_{k \in \cl{M}_{l,n} \rmv\backslash \{m\}} \!\! \phi^{(p-1)}_{k \rightarrow l}(\bd{x}_{l,n}) \ist\ist. \label{eq:extr_infor_1}\\[-5.5mm]
&\nonumber
\end{align}
This recursion is initialized with $b^{(0)}(\bd{x}_{l,n}) = \phi_{\rightarrow n}(\bd{x}_{l,n})$, 
$\psi^{(0)}_{m \rightarrow l}(\bd{x}_{m,n}) = \phi_{\rightarrow n}(\bd{x}_{m,n})$, and $\psi^{(0)}_{l \rightarrow m}(\bd{x}_{l,n}) = \phi_{\rightarrow n}(\bd{x}_{l,n})$.
The messages and beliefs involved in calculating $b^{(p)}(\bd{x}_{l,n})$ and $b^{(p)}(\bd{x}_{m,n})$ are shown in Fig.\ \ref{fig:coslat_fg}. 
We note that messages entering or leaving an object variable node in Fig.\ \ref{fig:coslat_fg} do not imply that there occurs any communication involving objects.

\vspace{-1.5mm}

\subsection{Discussion}
\label{sec:messpass:discuss}

According to \eqref{eq:marg_CoSLAT_agent} and \eqref{eq:marg_CoSLAT_T}, the agent beliefs $b^{(p)}(\bd{x}_{l,n})$ and object beliefs $b^{(p)}(\bd{x}_{m,n})$
involve the product of all the messages passed to the corresponding variable node $l$ and $m$, respectively. 
Similarly, according to \eqref{eq:extr_infor_2} and \eqref{eq:extr_infor_1}, the extrinsic informations $\psi^{(p-1)}_{m \rightarrow l}(\bd{x}_{m,n})$ and 
$\psi^{(p-1)}_{l \rightarrow m}(\bd{x}_{l,n})$ involve the product of all the messages passed to the corresponding variable node $m$ and $l$, 
respectively, except the message of the receiving factor node $f(\bd{y}_{l,m;n} | \bd{x}_{l,n}\ist,\bd{x}_{m,n})$.
Furthermore, according to \eqref{eq:meas_mess_1}, the extrinsic information $\psi^{(p-1)}_{k \rightarrow l}(\bd{x}_{k,n})$
passed from variable node $\bd{x}_{k,n}$ to factor node $f(\bd{y}_{k,l;n} | \bd{x}_{k,n}\ist,\bd{x}_{l,n})$ 
is used for calculating the message $\phi_{k \rightarrow l}^{(p)}(\bd{x}_{l,n})$ passed from that factor node to the respective other adjacent variable node $\bd{x}_{l,n}$. 
A similar discussion applies to \eqref{eq:meas_mess_3} and $\psi^{(p-1)}_{l \rightarrow m}(\bd{x}_{l,n})$. 

Two remarks are in order.  First, for low complexity, communication requirements, and latency, 
\pagebreak 
messages are sent only forward in time and iterative message passing 
is performed for each time individually. As a consequence, the message (extrinsic information) from 
variable node $\bd{x}_{k,n-1}$ to factor node $f(\bd{x}_{k,n} | \bd{x}_{k,n-1})$ equals the belief $b^{(P)}(\bd{x}_{k,n-1})$ (see \eqref{eq:pred_mess}), 
and $\phi_{\rightarrow n}(\bd{x}_{k,n})$ in \eqref{eq:pred_mess} (for $n$ fixed) remains unchanged during all message passing iterations.
Second, for any $k \in \AA$, as no information from the factor node $f(\bd{y}_{l,k;n} | \bd{x}_{l,n},\bd{x}_{k,n})$ is used in the calculation of $b^{(p-1)}(\bd{x}_{k,n})$ according to \eqref{eq:marg_CoSLAT_agent} and \eqref{eq:meas_mess_1}, $b^{(p-1)}(\bd{x}_{k,n})$
is used in \eqref{eq:meas_mess_1} as the extrinsic information passed to the factor node $f(\bd{y}_{l,k;n} | \bd{x}_{l,n},\bd{x}_{k,n})$. 
A similar message computation order is used in the SPAWN algorithm for CS \cite{wymeersch,lien}. This order significantly reduces the computational complexity since it avoids the computation of extrinsic informations exchanged among agent variable nodes. It also reduces the amount of communication between 
agents because beliefs passed between agents can be broadcast, whereas the exchange of extrinsic information would require separate 
point-to-point communications between agents \cite{wymeersch,lien}. 

Contrary to classical sequential Bayesian filtering \cite{doucet}, which only exploits the temporal conditional independence structure of the estimation
problem, the proposed BP scheme \eqref{eq:marg_CoSLAT_agent}--\eqref{eq:extr_infor_1} also exploits the spatial conditional 
independency structure. In fact, increasing the number of agents or objects leads to additional variable nodes in the factor graph 
\emph{but not to a higher dimension of the messages passed between the nodes}. As a consequence, the 
computational complexity scales very well in the numbers of agents and objects. As verified
in Section \ref{sec:compParticle}, a comparable scaling behavior cannot be achieved with classical Bayesian filtering techniques.

\begin{figure}
\vspace{.5mm}
\centering
\psfrag{L1}[l][l][0.8]{\raisebox{0mm}{\hspace{.5mm}CS}}
\psfrag{L2}[l][l][0.8]{\raisebox{0.3mm}{\hspace{.5mm}DT}}
\psfrag{L3}[l][l][0.8]{\raisebox{0mm}{\hspace{-.7mm}CS}}
\psfrag{L4}[l][l][0.8]{\raisebox{0.3mm}{\hspace{-.7mm}DT}}
\psfrag{S1}[l][l][0.8]{\raisebox{-5mm}{\hspace{-5.5mm}$b^{(P)}_{\AA,n-1}$}}
\psfrag{S2}[l][l][0.8]{\raisebox{-6mm}{\hspace{-5.5mm}$b^{(P)}_{\AA,n-1}$}}
\psfrag{S3}[l][l][0.8]{\raisebox{-5mm}{\hspace{-5.5mm}$b^{(P)}_{\OO,n-1}$}}
\psfrag{S4}[l][l][0.8]{\raisebox{-5mm}{\hspace{-5.5mm}$b^{(P)}_{\OO,n-1}$}}
\psfrag{S5}[l][l][0.8]{\raisebox{-5mm}{\hspace{-2mm}$b^{(P)}_{\AA,n}$}}
\psfrag{S6}[l][l][0.8]{\raisebox{-5mm}{\hspace{-2mm}$b^{(P)}_{\OO,n}$}}
\psfrag{S7}[l][l][0.8]{\raisebox{-6mm}{\hspace{-2mm}$b^{(P)}_{\AA,n}$}}
\psfrag{S8}[l][l][0.8]{\raisebox{-5mm}{\hspace{-2mm}$b^{(P)}_{\OO,n}$}}
\psfrag{S9}[l][l][0.8]{\raisebox{-4mm}{\hspace{0mm}$\hat{\bd{x}}_{\AA,n}^{(P)}$}}
\psfrag{S10}[l][l][0.8]{\raisebox{-4mm}{\hspace{-3mm}$\psi_{\AA \rightarrow \OO,n}^{(p)}$}}
\psfrag{S11}[l][l][0.8]{\raisebox{-4mm}{\hspace{-3mm}$\psi_{\OO \rightarrow \AA,n}^{(p)}$}}
\psfrag{A1}[l][l][0.8]{\raisebox{-4mm}{\hspace{0mm}(a)}}
\psfrag{A2}[l][l][0.8]{\raisebox{-4mm}{\hspace{0mm}(b)}}
\includegraphics[scale=.7]{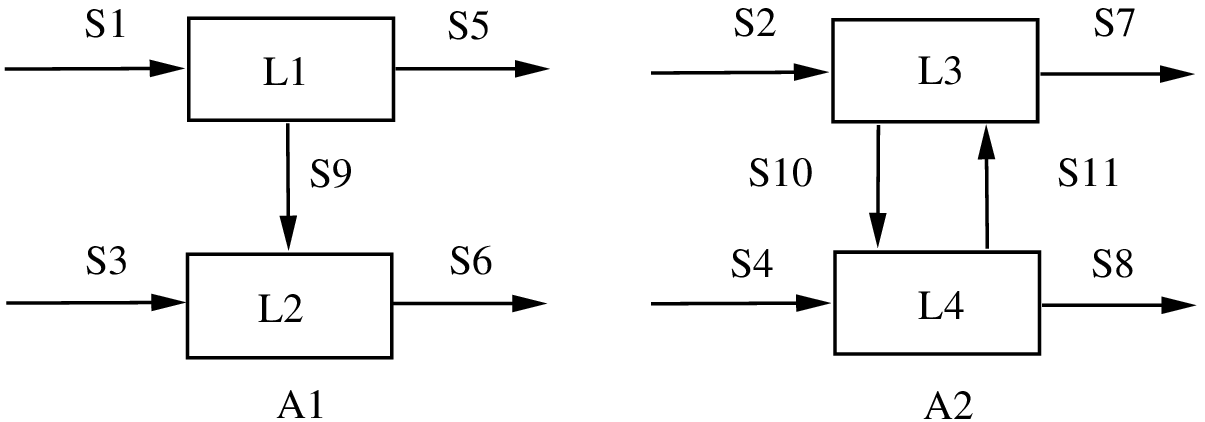}
\vspace{-1mm}
\caption{Block diagram of (a) separate CS and DT and (b) the proposed scheme,
with $b^{(P)}_{\AA,n} \rmv\triangleq\rmv \big\{ b^{(P)}(\bd{x}_{l,n}) \big\}_{l \in \AA}\ist$, 
$b^{(P)}_{\OO,n} \rmv\triangleq\rmv \big\{ b^{(P)}(\bd{x}_{m,n}) \big\}_{m \in \OO}\ist$, 
$\psi_{\AA \rightarrow \OO,n}^{(p)} \!\!\triangleq \big\{ \psi^{(p)}_{l \rightarrow m}(\bd{x}_{l,n})\big\}_{l \in \AA_{m,n},\ist m \in \OO}$, 
$\psi_{\OO \rightarrow \AA,n}^{(p)} \!\!\triangleq\rmv \big\{ \psi^{(p)}_{m \rightarrow l}(\bd{x}_{m,n})\big\}_{m \in \cl{M}^{\OO}_{l,n},\ist l \in \AA}$, 
and $\hat{\bd{x}}_{\AA,n}^{(P)} \!\triangleq\! \big( \hat{\bd{x}}_{l,n}^{(P)}\big)_{l \in \AA}$. In separate CS and DT, the final agent state estimates 
$\hat{\bd{x}}_{\AA,n}^{(P)}$ are transferred from CS to DT. In the proposed scheme, probabilistic information (the extrinsic informations 
$\psi_{\AA \rightarrow \OO,n}^{(p)}\rmv$ and $\psi_{\OO \rightarrow \AA,n}^{(p)}$) is transferred between CS and DT in each message passing iteration $p$.}
\label{fig:turbo}
\vspace{0mm}
\end{figure}

The factor graph in Fig.\ \ref{fig:coslat_fg} and the corresponding BP scheme \eqref{eq:marg_CoSLAT_agent}--\eqref{eq:extr_infor_1}
combine CS and DT into a unified, coherent estimation technique. Indeed, in contrast to the conventional approach of separate CS and DT---i.e.,
first performing CS to localize the agents and then, based on the estimated agent locations, performing DT to localize the objects---our BP scheme
exchanges probabilistic information between the CS and DT parts of the factor graph. Thereby,
uncertainties in one stage are taken into account by the respective other stage, and the performance of both stages can be improved.
This information transfer, which will be further discussed in Sections \ref{sec:object_belief} and \ref{sec:agent_beliefs}, 
is the main reason for the superior performance of the proposed joint CS-DT algorithm; it is visualized and contrasted with the conventional approach 
in Fig.\ \ref{fig:turbo}.

\vspace{-1.5mm}

\section{Particle-Based Processing} 
\label{sec:NBP_CoSLAT}

\vspace{.5mm}

Because of the nonlinear and non-Gaussian state transition model \eqref{eq_statrans} and measurement model \eqref{eq:meas_mod}, 
a closed-form evaluation of the integrals and message products in \eqref{eq:marg_CoSLAT_agent}--\eqref{eq:extr_infor_1} is typically impossible.  
Therefore, we next present a new low-complexity particle-based implementation of the BP scheme \eqref{eq:marg_CoSLAT_agent}--\eqref{eq:extr_infor_1}. 
This implementation uses particle representations (PRs) of beliefs and messages, which approximate distributions in terms of randomly drawn particles (samples)
$\bd{x}^{(j)}$ and weights $w^{(j)}\rmv$, for $j \in \{1,\ldots,J\}$. As in conventional nonparametric BP \cite{ihler,lien}, the 
operations of \emph{message filtering} and \emph{message multiplication} are performed.  Message filtering 
calculates the prediction message \eqref{eq:pred_mess} and equals the message filtering operation of nonparametric BP. However, for message multiplication 
in \eqref{eq:marg_CoSLAT_agent}, \eqref{eq:marg_CoSLAT_T}, \eqref{eq:extr_infor_2}, and \eqref{eq:extr_infor_1}, 
the ``stacking technique'' introduced in \cite{meyer14sigma} is used. This technique avoids an explicit calculation of the measurement messages 
\eqref{eq:meas_mess_1} and \eqref{eq:meas_mess_3} and does not use computationally intensive kernel density estimates, which are required 
by nonparametric BP. Thereby, its complexity is only linear in the number of particles.
We note that an alternative message multiplication scheme that also avoids the use of kernel density estimates and whose complexity is linear 
in the number of particles was proposed in \cite{briers05}. This scheme constructs an approximate proposal distribution in order to calculate 
weighted particles for beliefs and messages. Our approach is different in that the proposal distribution is formed simply by ``stacking'' incoming beliefs, 
and the calculation of particles and weights for incoming messages is avoided (see Section \ref{sec:NBPlowcomplexity}).

\vspace{-1mm}

\subsection{Message Filtering} \label{sec:messfilter} 

The message filtering operation reviewed in the following is analogous to the prediction step of the 
sampling importance resampling particle filter \cite{arulampalam}. Particle-based calculation of \eqref{eq:pred_mess} means that we obtain
a PR $\big\{\big(\bd{x}_{k,n}^{(j)}, w_{k,n}^{(j)} \big) \big\}_{j=1}^{J}$ of 
$\phi_{\rightarrow n}(\bd{x}_{k,n}) =\! \int \rmv\rmv f(\bd{x}_{k,n}|\bd{x}_{k,n-1})\, b^{(P)}(\bd{x}_{k,n-1}) \ist d\bd{x}_{k,n-1}$ 
from a PR $\big\{ \big(\bd{x}_{k,n-1}^{(j)} ,w_{k,n-1}^{(j)} \big) \big\}_{j=1}^{J}$ of $b^{(P)}(\bd{x}_{k,n-1})$. 
This can be easily done by recognizing that the above integral is a marginalization of $f(\bd{x}_{k,n}|\bd{x}_{k,n-1}) \, b^{(P)}(\bd{x}_{k,n-1})$.
Motivated by this interpretation, we first establish a PR $\big\{ \big( \bd{x}_{k,n}^{(j)} , \bd{x}_{k,n-1}^{(j)},w_{k,n-1}^{(j)} \big) \big\}_{j=1}^{J}$ of 
$f(\bd{x}_{k,n}|\bd{x}_{k,n-1}) \ist b^{(P)}(\bd{x}_{k,n-1})$ by drawing for each particle $\big(\bd{x}_{k,n-1}^{(j)} ,w_{k,n-1}^{(j)} \big)$ 
representing $b^{(P)}(\bd{x}_{k,n-1})$ one particle $\bd{x}_{k,n}^{(j)}$ from $f(\bd{x}_{k,n}|\bd{x}^{(j)}_{k,n-1})$.
Then, removing $\big\{ \bd{x}_{k,n-1}^{(j)} \big\}_{j=1}^{J}$ from $\big\{\big(\bd{x}_{k,n}^{(j)} ,\bd{x}_{k,n-1}^{(j)} , w_{k,n-1}^{(j)} \big) \big\}_{j=1}^{J}$ 
is the Monte Carlo implementation of the above marginalization \cite{ristic}.
This means that $\big\{\big(\bd{x}_{k,n}^{(j)} ,w_{k,n}^{(j)} \big) \big\}_{j=1}^{J}$ with $w_{k,n}^{(j)} = w_{k,n-1}^{(j)}$ for all 
$j \rmv\in\rmv \{1,\dots,J\}$ constitutes the desired PR of $\phi_{\rightarrow n}(\bd{x}_{k,n})$.

\vspace{-1mm}

\subsection{Message Multiplication} \label{sec:NBPlowcomplexity}

Next, we propose a message multiplication scheme for calculating the beliefs in \eqref{eq:marg_CoSLAT_agent} and \eqref{eq:marg_CoSLAT_T} 
and the extrinsic informations in \eqref{eq:extr_infor_2} and \eqref{eq:extr_infor_1}. For concreteness, we present
the calculation of the agent beliefs \eqref{eq:marg_CoSLAT_agent}; the object beliefs \eqref{eq:marg_CoSLAT_T} and extrinsic informations \eqref{eq:extr_infor_2} 
and \eqref{eq:extr_infor_1} are calculated in a similar manner. 

Following \cite{meyer14sigma}, we consider the ``stacked state'' $\bar{\bd{x}}_{l,n} \!\triangleq \big(\bd{x}_{k,n}\big)_{k \in\ist \{l\} \cup \cl{M}_{l,n}}\rmv$, 
which consists of the agent state $\bd{x}_{l,n}$ and the states $\bd{x}_{k,n}$ of all measurement partners $k \in\cl{M}_{l,n}$ of agent $l$. 
Using \eqref{eq:meas_mess_1} in \eqref{eq:marg_CoSLAT_agent}, one readily obtains
\be
\label{eq:b_alt}
b^{(p)}(\bd{x}_{l,n}) = \int b^{(p)}(\bar{\bd{x}}_{l,n}) \, d\bar{\bd{x}}_{l,n}^{\sim l} \,,
\vspace{-2mm}
\ee
where 
\begin{align}
\hspace{-1.5mm}b^{(p)}(\bar{\bd{x}}_{l,n}) &\propto \phi_{\rightarrow n}(\bd{x}_{l,n}) \nn\\[.7mm]
  &\hspace{4mm}\times\!\!\rmv \prod_{l' \in \cl{M}^{\AA}_{l,n}} \!\! f(\bd{y}_{l,l';n} | \bd{x}_{l,n},\bd{x}_{l'\!,n}) \,b^{(p-1)}(\bd{x}_{l'\!,n}) \nn \\
  &\hspace{4mm}\times\!\!\! \prod_{m \in \cl{M}^{\OO}_{l,n}} 
  \!\!\rmv f(\bd{y}_{l,m;n} | \bd{x}_{l,n},\bd{x}_{m,n}) \,\psi^{(p-1)}_{m \rightarrow l}(\bd{x}_{m,n}) \label{eq:b_alt_stacked}\\[-6mm]
\nn
\end{align}
and $d\bar{\bd{x}}_{l,n}^{\sim l} \triangleq \prod_{k \in \cl{M}_{l,n}} \! d\bd{x}_{k,n}$. To obtain a PR of $b^{(p)}(\bd{x}_{l,n})$, first a PR 
$\big\{ \big( \bar{\bd{x}}_{l,n}^{(j)}, w_{l,n}^{(j)} \big) \big\}_{j=1}^{J}$ of $b^{(p)}(\bar{\bd{x}}_{l,n})$ is calculated as explained presently.
Then,  $\big\{ \big( \bd{x}_{l,n}^{(j)}, w_{l,n}^{(j)} \big) \big\}_{j=1}^{J}$ is a PR of $b^{(p)}(\bd{x}_{l,n})$. 
This is because $\bd{x}_{l,n}^{(j)}$, as a subvector of $\bar{\bd{x}}^{(j)}_{l,n}$, can be obtained by removing from $\bar{\bd{x}}^{(j)}_{l,n}$ 
the other subvectors $\bd{x}_{k,n}^{(j)}$, $k \in \cl{M}_{l,n}$, which is the Monte Carlo implementation of the marginalization \eqref{eq:b_alt}
(cf.\ Section \ref{sec:messfilter}). Finally, a resampling \cite{ristic} produces equally weighted particles representing $b^{(p)}(\bd{x}_{l,n})$. 

A PR $\big\{ \big( \bar{\bd{x}}_{l,n}^{(j)}, w_{l,n}^{(j)} \big) \big\}_{j=1}^{J}$ of $b^{(p)}(\bar{\bd{x}}_{l,n})$ can be calculated 
via importance sampling using the proposal 
\vspace{1mm}
distribution\footnote{An 
alternative proposal distribution that is more appropriate if agent $l$ is static or if the prediction message is very noninformative is presented in Section \ref{sec:CoSLAT_inital}.} 
\begin{align}
q(\bar{\bd{x}}_{l,n}) \,\triangleq\, \phi_{\rightarrow n}(\bd{x}_{l,n}) \rrmv\rmv \prod_{l' \in \cl{M}^{\AA}_{l,n}} \!\! b^{(p-1)}(\bd{x}_{l'\!,n}) \rrmv\rmv 
  \prod_{m \in \cl{M}^{\OO}_{l,n}} \!\!\rmv \psi^{(p-1)}_{m \rightarrow l}(\bd{x}_{m,n}) \,. \nn\\[-3.5mm]
\label{eq:proposal}\\[-7mm]
\nn
\end{align}
There is no need to draw particles $\big\{\bar{\bd{x}}^{(j)}_{l,n}\big\}_{j=1}^{J}$ from $q(\bar{\bd{x}}_{l,n})$ because such particles can be obtained simply by 
stacking particles $\big\{ \bd{x}_{l,n}^{(j)} \big\}_{j=1}^J$ representing $\phi_{\rightarrow n}(\bd{x}_{l,n})$,
particles $\big\{ \bd{x}_{l'\!,n}^{(j)} \big\}_{j=1}^J$ representing $b^{(p-1)}(\bd{x}_{l'\!,n})$, $l' \!\in\rmv \cl{M}^{\AA}_{l,n}$, 
and particles $\big\{ \bd{x}_{m,n}^{(j)} \big\}_{j=1}^J$ representing $\psi^{(p-1)}_{m \rightarrow l}(\bd{x}_{m,n})$, $m \!\in\rmv \cl{M}^{\OO}_{l,n}$.
(Particles representing $\phi_{\rightarrow n}(\bd{x}_{l,n})$ were obtained by message filtering. The other particles, for $p \!\ge\! 2$, were 
calculated at iteration $p \!-\! 1$. For $p \!=\! 1$, since $b^{(0)}(\bd{x}_{l,n}) = \phi_{\rightarrow n}(\bd{x}_{l,n})$, $\psi^{(0)}_{m \rightarrow l}(\bd{x}_{m,n}) = \phi_{\rightarrow n}(\bd{x}_{m,n})$, and $\psi^{(0)}_{l \rightarrow m}(\bd{x}_{l,n}) = \phi_{\rightarrow n}(\bd{x}_{l,n})$, these particles are identical to those obtained by message filtering.) 
Weights $w_{l,n}^{(j)}$ corresponding to the stacked particles $\big\{\bar{\bd{x}}^{(j)}_{l,n}\big\}_{j=1}^{J}$ are then obtained by calculating 
$\tilde{w}_{l,n}^{(j)} \propto b^{(p)}(\bar{\bd{x}}^{(j)}_{l,n})/q(\bar{\bd{x}}^{(j)}_{l,n})$ followed by a normalization. Using \eqref{eq:b_alt_stacked}, 
the nonnormalized weights are obtained 
\vspace{1mm}
as
\[
\tilde{w}^{(j)}_{l,n} =\rmv\rmv\rrmv \prod_{l' \in \cl{M}^{\AA}_{l,n}} \rrmv\rmv\rmv f(\bd{y}_{l,l';n}|\bd{x}^{(j)}_{l,n}, \bd{x}^{(j)}_{l'\!,n}) \rrmv\rmv\rmv 
  \prod_{m \in \cl{M}^{\OO}_{l,n}} \rrmv\rmv\rmv f(\bd{y}_{l,m;n}|\bd{x}^{(j)}_{l,n}, \bd{x}^{(j)}_{m,n})\,.
\]

This algorithm avoids kernel density estimation, which is required by conventional nonparametric BP \cite{ihler,lien}. 
Its complexity scales as $\cl{O}\big(|\cl{M}_{l,n}| \ist J\big)$, i.e., 
\pagebreak 
only linearly in the number of particles $J$. The dimension of the distribution 
$b^{(p)}(\bar{\bd{x}}_{l,n})$ involved in the importance sampling scheme is $|\cl{M}_{l,n}| + 1$, and thus typically considerably higher than that of the beliefs
$b^{(p)}(\bd{x}_{l,n})$ involved in the importance sampling scheme of nonparametric BP \cite{ihler,lien}. Nevertheless, we will see in Section \ref{sec:simres} 
that, if the number of neighbors $|\cl{M}_{l,n}|$ is not too large, the number of particles $J$ required for high accuracy
is not larger than for nonparametric BP.

\vspace{-1.5mm}

\subsection{Estimation} \label{sec:estimation} 

The particle-based BP algorithm described above produces PRs $\big\{ \big( \bd{x}_{k,n}^{(j)}, w_{k,n}^{(j)} \big) \big\}_{j=1}^{J}\rmv$ of 
the state beliefs $b^{(p)}(\bd{x}_{k,n})$, $k \!\in\! \EE$.\linebreak 
An approximation of the estimate $\hat{\bd{x}}^{\text{MMSE}}_{k,n}$ in \eqref{eq:mmse_slat} is then obtained from the respective PR 
\vspace{-.5mm}
as
\be
\hat{\bd{x}}_{k,n} \ist=\ist \sum_{j=1}^{J} w_{k,n}^{(j)} \bd{x}_{k,n}^{(j)} \,. 
\label{eq:et_part}
\vspace{-.5mm}
\ee

\section{Distributed Algorithm}
\label{sec:slat_algor}

\vspace{.5mm}

We next develop a distributed algorithm that combines the particle-based BP algorithm discussed in Section \ref{sec:NBP_CoSLAT}
with a consensus scheme \cite{farahmand11}. The overall organization of this algorithm is as follows. Each agent $l \rmv\in\rmv \AA$ performs 
particle-based estimation of its own state $\bd{x}_{l,n}$ and of the states $\bd{x}_{m,n}$, $m \rmv\in\rmv \OO$ of all the objects. Thus, the 
calculations required to estimate an agent state $\bd{x}_{l,n}$, $l \rmv\in\rmv \AA$ are performed only once in the network (at agent $l$), 
whereas certain calculations required to estimate an object state $\bd{x}_{m,n}$, $m \rmv\in\rmv \OO$ are performed $|\AA|$ times 
(at each agent $l \rmv\in\rmv \AA$). Accordingly, each agent belief $b^{(p)}(\bd{x}_{l,n})$ is stored (temporarily, i.e., during one message passing iteration) 
only at the respective agent $l$ whereas copies of all object beliefs $b^{(p)}(\bd{x}_{m,n})$ are stored (temporarily) at all agents $l \in \AA$. 
However, all the calculations performed at any given agent $l \rmv\in\rmv \AA$ are collaborative in that they use probabilistic information 
related to all the other agents and objects. The proposed distributed algorithm requires only communication between neighboring agents 
to disseminate this probabilistic information. The distributed calculation of the object beliefs, agent beliefs, and extrinsic informations will be discussed in 
the next three subsections.

\vspace{-1mm}

\subsection{Distributed Calculation of the Object Beliefs}
\label{sec:object_belief}

\vspace{.5mm}

Estimation of the object states $\bd{x}_{m,n}$, $m \rmv\in\rmv \OO$ from $\bd{y}_{1:n}$ according to \eqref{eq:et_part} 
essentially amounts to a particle-based computation of $b^{(p)}(\bd{x}_{m,n})$. The following discussion describes the calculations associated with the red dotted box in Fig.\ \ref{fig:coslat_fg}. According to \eqref{eq:marg_CoSLAT_T} and \eqref{eq:pred_mess}, the object belief $b^{(p)}(\bd{x}_{m,n})$, $p \rmv\in\rmv \{1,\dots,P\}$ approximating $f(\bd{x}_{m,n}|\bd{y}_{1:n})$ is given 
\vspace{.8mm}
by
\be
b^{(p)}(\bd{x}_{m,n}) \,\propto\, \phi_{\rightarrow n}(\bd{x}_{m,n}) \, \Phi^{(p)}_{m,n}(\bd{x}_{m,n}) \,,
\label{eq:marg_1}
\vspace{-1mm}
\ee
with 
\begin{align}
\hspace{-1mm}\phi_{\rightarrow n}(\bd{x}_{m,n}) &\,= \int \rmv f(\bd{x}_{m,n}|\bd{x}_{m,n-1}) \, b^{(P)}(\bd{x}_{m,n-1}) \, d\bd{x}_{m,n-1} 
  \nonumber\\[-1.5mm]
&\label{eq:marg_01}\\[-9mm]
\nonumber
\end{align}
and
\vspace{1.5mm}
\begin{equation}
\label{eq:M_prod}
\Phi_{m,n}^{(p)}(\bd{x}_{m,n}) \,\triangleq\rmv \prod_{l \in \AA_{m,n}} \!\! \phi^{(p)}_{l \rightarrow m}(\bd{x}_{m,n}) \,.
\end{equation}
According to \eqref{eq:marg_1}, each agent has to calculate the prediction message $\phi_{\rightarrow n}(\bd{x}_{m,n})$
in \eqref{eq:marg_01} and the measurement message product $\Phi_{m,n}^{(p)}(\bd{x}_{m,n})$ in \eqref{eq:M_prod}. The messages 
$\phi^{(p)}_{l \rightarrow m}(\bd{x}_{m,n})$ contained in \eqref{eq:M_prod} involve the extrinsic informations $\psi^{(p-1)}_{l \rightarrow m}(\bd{x}_{l,n})$ 
(see \eqref{eq:meas_mess_3}); particle-based calculation of the latter will be discussed in Section \ref{sec:extrinfo}. However,
at each agent at most one message $\phi^{(p)}_{l \rightarrow m}(\bd{x}_{m,n})$ is available (for a given $m$). We will solve this problem by 
means of a consensus scheme.

\vspace{1.5mm}

\subsubsection{Particle-based Calculation of $b^{(p)}(\bd{x}_{m,n})$}
\label{sec:calc_belief_m}

An approximate particle-based calculation of $b^{(p)}(\bd{x}_{m,n}) \propto \phi_{\rightarrow n}(\bd{x}_{m,n}) \, \Phi^{(p)}_{m,n}(\bd{x}_{m,n})$ 
in \eqref{eq:marg_1} can be obtained via importance sampling with proposal distribution $\phi_{\rightarrow n}(\bd{x}_{m,n})$.
First, based on \eqref{eq:marg_01}, particles $\big \{\bd{x}_{m,n}^{(j)}\big \}_{j=1}^{J}$ representing $\phi_{\rightarrow n}(\bd{x}_{m,n})$ are calculated from 
particles representing $b^{(P)}(\bd{x}_{m,n-1})$ by means of message filtering (cf.\ Section \ref{sec:messfilter}; note that particles representing $b^{(P)}(\bd{x}_{m,n-1})$ 
were calculated by each agent at time $n \!-\! 1$). Next, weights $\big\{w_{m,n}^{(j)}\big\}_{j=1}^{J}$ are calculated 
\vspace{-.3mm}
as
\be
\tilde{w}_{m,n}^{(j)} \ist=\, \Phi_{m,n}^{(p)}(\bd{x}^{(j)}_{m,n})
\label{eq:weight_m}
\vspace{.3mm}
\ee
followed by a normalization. Finally, resampling is performed to obtain equally weighted particles representing $b^{(p)}(\bd{x}_{m,n})$. However, this particle-based 
implementation presupposes that the message product $\Phi_{m,n}^{(p)}(\bd{x}_{m,n})$ evaluated at the particles $\big \{\bd{x}_{m,n}^{(j)}\big \}_{j=1}^{J}$
is available at the agents. 

\vspace{.5mm}

\subsubsection{Distributed Evaluation of $\Phi^{(p)}_{m,n}(\cdot)$}
\label{sec:calcobjectmarg}

For a distributed computation of $\Phi_{m,n}^{(p)}(\bd{x}_{m,n}^{(j)})$, $j \in \{1,\dots,J\}$, we first note that \eqref{eq:M_prod} 
for $\bd{x}_{m,n} = \bd{x}_{m,n}^{(j)}$ can be written as
\be
\Phi_{m,n}^{(p)}(\bd{x}_{m,n}^{(j)}) \,=\, \exp\rmv \big( |\AA| \ist \chi_{m,n}^{(p,j)} \big) \,,
\label{eq:M_prod_log}
\vspace{-1.5mm}
\ee
with
\begin{align}
&\chi_{m,n}^{(p,j)} \,\triangleq\ist \frac{1}{|\AA|} \sum_{l \in \AA_{m,n}} \!\! \log \, \phi^{(p)}_{l \rightarrow m}(\bd{x}_{m,n}^{(j)}) \,, \quad j \in \{1,\dots,J\} \,. \nn\\[-4mm]
&\label{eq:M_prod_log_chi}\\[-7mm]
\nn
\end{align}
Thus, $\Phi_{m,n}^{(p)}(\bd{x}_{m,n}^{(j)})$ is expressed in terms of the arithmetic average $\chi_{m,n}^{(p,j)}$. For each $j$, following the ``consensus--over--weights'' 
approach of \cite{farahmand11}, this average can be computed in a distributed manner by a consensus or gossip scheme \cite{olfati07,dimakis10},
in which each agent communicates only with neighboring agents. 
These schemes are iterative; in each iteration $i$, they compute an internal state $\zeta_{l,m;n}^{(j,i)}$ at each agent $l$. This internal state is initialized as
\begin{equation}
\zeta_{l,m;n}^{(j,0)} = \begin{cases} 
     \log \, \phi_{l \rightarrow m}^{(p)}(\bd{x}^{(j)}_{m,n}) \ist, & \quad l \in \AA_{m,n}\\[.1mm]
     0 & \quad l \notin \AA_{m,n}\ist.
   \end{cases}
\label{eq:CoSlatInit}
\vspace{-1.5mm}
\end{equation}
Here, $\phi_{l \rightarrow m}^{(p)}(\bd{x}^{(j)}_{m,n})$ is computed by means of a Monte Carlo approximation \cite{doucet} of the integral in \eqref{eq:meas_mess_3}, i.e., 
\be
\phi_{l \rightarrow m}^{(p)}(\bd{x}^{(j)}_{m,n}) \ist\approx\ist \frac{1}{J} \sum_{j'=1}^{J} \rmv f(\bd{y}_{l,m;n} | \bd{x}^{(j')}_{l,n}\!,\bd{x}^{(j)}_{m,n}) \,.
\label{eq:phi_approx}
\vspace{-.5mm}
\ee
This uses the particles $\big\{\bd{x}^{(j)}_{l,n}\big\}^{J}_{j=1}$ representing $\psi^{(p-1)}_{l \rightarrow m}(\bd{x}_{l,n})$, 
whose calculation will be discussed in Section \ref{sec:extrinfo}. If---as assumed in Section \ref{sec:sceass:systmod}---the 
communication graph is connected, then for $i \rmv\to\rmv \infty$ the internal state 
\pagebreak 
$\zeta_{l,m;n}^{(j,i)}$ converges to the average $\chi_{m,n}^{(p,j)}$ 
in \eqref{eq:M_prod_log_chi} \cite{olfati07,dimakis10} (more precisely, to an approximation of $\chi_{m,n}^{(p,j)}$, due to the approximation \eqref{eq:phi_approx}). 
Thus, for a sufficiently large number $C$ of iterations $i$, because of \eqref{eq:M_prod_log}, 
a good approximation of $\Phi_{m,n}^{(p)}(\bd{x}^{(j)}_{m,n})$ is obtained at each agent by 
\be
\Phi_{m,n}^{(p)}(\bd{x}^{(j)}_{m,n}) \,\approx\, \exp \rmv\big(|\AA| \ist \zeta_{l,m;n}^{(j,C)}\big) \,.
\label{eq:Phi_consensus}
\ee
Here, the number of agents $|\AA|$ can be determined in a distributed way by using another consensus or gossip algorithm at time $n=0$ \cite{pham2009}. 
Furthermore, an additional max-consensus scheme has to be used to obtain perfect consensus on the weights $\tilde{w}_{m,n}^{(j)}$ 
in \eqref{eq:weight_m} and, in turn, identical particles at all agents \cite{farahmand11}. The max-consensus converges in $I$ iterations, where $I$ is the 
diameter of the communication graph \cite{olfatisaber}. Finally, the pseudo-random number generators of all agents (which are used for drawing particles) 
have to be synchronized, i.e., initialized with the same seed at time $n=0$. This distributed evaluation of $\Phi_{m,n}^{(p)}(\bd{x}^{(j)}_{m,n})$ 
requires only local communication: in each iteration, for each of the $J$ instances of the consensus or gossip scheme, $J$ real values are broadcast 
by each agent to neighboring agents \cite{olfati07,dimakis10}. This holds for averaging and maximization separately.

As an alternative to this scheme, the likelihood consensus scheme \cite{hlinka14adaptation, hlinkaMag13} can be employed to provide an approximation 
of the functional form of $\Phi_{m,n}^{(p)}(\bd{x}_{m,n})$ to each agent, again using only local communication \cite{meyer12}. 
The likelihood consensus scheme does not require additional max-consensus algorithms and synchronized pseudo-random number generators, 
but tends to require a more informative proposal distribution for message multiplication (cf.\ Section \ref{sec:NBPlowcomplexity}).

\vspace{1.5mm}

\subsubsection{Probabilistic Information Transfer}
According to \eqref{eq:meas_mess_3}, the messages $\phi^{(p)}_{l \rightarrow m}(\bd{x}_{m,n})$ occurring in 
$\Phi_{m,n}^{(p)}(\bd{x}_{m,n}) = \prod_{l \in \AA_{m,n}} \!\! \phi^{(p)}_{l \rightarrow m}(\bd{x}_{m,n})$ (see \eqref{eq:M_prod})
involve the extrinsic informations $\psi^{(p-1)}_{l \rightarrow m}(\bd{x}_{l,n})$ of all agents $l$ observing object $m$, i.e., $l \rmv\in\rmv \AA_{m,n}$. 
Therefore, they constitute an information transfer from the CS part of the algorithm to the DT part 
(cf.\ Fig.\ \ref{fig:turbo}(b) and, in more detail, the directed edges entering the red dotted box in Fig.\ \ref{fig:coslat_fg}). 
The estimation of object state $\bd{x}_{m,n}$ is based on the belief $b^{(p)}(\bd{x}_{m,n})$ as given by \eqref{eq:marg_1}, and thus on 
$\Phi_{m,n}^{(p)}(\bd{x}_{m,n})$. This improves on pure DT because probabilistic information about the states of the agents 
$l \rmv\in\rmv \AA_{m,n}$---provided by $\psi^{(p-1)}_{l \rightarrow m}(\bd{x}_{l,n})$---is taken into account. 
By contrast, pure DT according to \cite{hlinka14adaptation,hlinkaMag13,farahmand11,savic14} uses the global likelihood function---involving the measurements of all agents---instead of 
$\Phi_{m,n}^{(p)}(\bd{x}_{m,n})$. This presupposes that the agent states are known. In separate CS and DT, estimates of the agent states 
provided by CS are used for DT, rather than probabilistic information about the agent states as is done in the proposed combined CS--DT algorithm. 
The improved accuracy of object state estimation achieved by our algorithm compared to separate CS and DT will be demonstrated in 
Section \ref{sec:statScenario}.

\vspace{-1mm}

\subsection{Distributed Calculation of the Agent Beliefs}
\label{sec:agent_beliefs}

\vspace{.5mm}

For a distributed calculation of the agent belief $b^{(p)}(\bd{x}_{l,n})$, $l \rmv\in\rmv \AA$, the following information is available at agent $l$:
(i) equally weighted particles representing $\psi^{(p-1)}_{m \rightarrow l}(\bd{x}_{m,n})$ for all objects $m \rmv\in\rmv \OO$ (whose calculation 
will be described in Section \ref{sec:extrinfo}); (ii) equally weighted particles representing $b^{(p-1)}(\bd{x}_{l'\!,n})$ for all neighboring agents 
$l' \!\in\rmv \cl{M}^{\AA}_{l,n}$ (which were received from these agents); and (iii) a PR of $b^{(P)}(\bd{x}_{l,n-1})$ (which was calculated at time $n \rmv-\! 1$).
Based on this information and the locally available measurements $\bd{y}_{l,k;n}$, $k \rmv\in\rmv \cl{M}_{l,n}$, a PR 
$\big\{ \big( \bd{x}_{l,n}^{(j)},w_{l,n}^{(j)} \big) \big\}_{j=1}^{J}$ of $b^{(p)}(\bd{x}_{l,n})$ can be calculated in a distributed manner by implementing 
\eqref{eq:marg_CoSLAT_agent}, using the particle-based message multiplication scheme presented in Section \ref{sec:NBPlowcomplexity}. 
Finally, resampling is performed to obtain equally weighted particles representing $b^{(p)}(\bd{x}_{l,n})$. This calculation of the agent beliefs improves 
on pure CS \cite{wymeersch} in that it uses the probabilistic information about the states of the objects $m \rmv\in\rmv \cl{M}^{\OO}_{l,n}$ provided by 
the messages $\psi^{(p-1)}_{m \rightarrow l}(\bd{x}_{m,n})$. This probabilistic information transfer from DT to CS is depicted 
in Fig.\ \ref{fig:turbo}(b) and, in more detail, by the directed edges leaving the red dotted box in Fig.\ \ref{fig:coslat_fg}. The resulting improved accuracy 
of agent state estimation will be demonstrated in Section \ref{sec:simres}. 

\vspace{-.5mm}

\subsection{Distributed Calculation of the Extrinsic Informations}
\label{sec:extrinfo}

\vspace{.5mm}

Since \eqref{eq:extr_infor_1} is analogous to \eqref{eq:marg_CoSLAT_agent} and \eqref{eq:extr_infor_2} is analogous to 
\eqref{eq:marg_CoSLAT_T}, particles for $\psi^{(p)}_{l \rightarrow m}(\bd{x}_{l,n})$ or $\psi^{(p)}_{m \rightarrow l}(\bd{x}_{m,n})$ can be calculated 
similarly as for the corresponding belief. However, in the case of $\psi^{(p)}_{m \rightarrow l}(\bd{x}_{m,n})$, the following shortcut reusing previous results 
can be used. According to \eqref{eq:marg_CoSLAT_T} and \eqref{eq:extr_infor_2}, 
$\psi_{m \rightarrow l}^{(p)}(\bd{x}_{m,n}) \propto b^{(p)}(\bd{x}_{m,n}) / \phi_{l \rightarrow m}^{(p)}(\bd{x}_{m,n})$.
Therefore, to obtain particles for $\psi^{(p)}_{m\rightarrow l}(\bd{x}_{m,n})$, 
we proceed as for $b^{(p)}(\bd{x}_{m,n})$ (see Sections \ref{sec:calc_belief_m} and \ref{sec:calcobjectmarg}) but replace 
$\exp \rmv\big(|\AA| \ist \zeta_{l,m;n}^{(j,C)}\big)$ in \eqref{eq:Phi_consensus} with $\exp\rmv \big( |\AA| \ist \zeta_{l,m;n}^{(j,C)} - \zeta_{l,m;n}^{(j,0)} \big)$.
Here, $\zeta_{l,m;n}^{(j,C)}$ and $\zeta_{l,m;n}^{(j,0)}$ are already available locally from the calculation of $b^{(p)}(\bd{x}_{m,n})$.

\vspace{-.5mm}

\subsection{Statement of the Distributed Algorithm}
\label{sec:CoSLAT_algor_statement}

\vspace{.5mm}

The proposed distributed CS--DT algorithm is obtained by combining the operations discussed in Sections \ref{sec:object_belief} through \ref{sec:extrinfo}, 
as summarized in the following.

\vspace*{3mm}
{
\hrule
\vspace{-.9mm}
\begin{center}
{\small\sc Algorithm 1:\, Distributed CS--DT Algorithm}
\vspace{-.7mm}
\end{center}
\hrule 
\vspace{2.5mm}

\small
\renewcommand{\baselinestretch}{1.2}\normalsize\small

\noindent
\emph{Initialization}:\, The recursive algorithm described below is initialized at time $n\!=\!0$ and agent $l$ 
with particles $\big\{ \bd{x}_{k,0}^{\prime (j)} \big\}_{j=1}^{J}$ drawn from a prior pdf $f(\bd{x}_{k,0})$, for $k \rmv\in\rmv \{l\} \cup \OO$.
\vspace{2mm}

\noindent
\emph{Recursion at time $n$}:\, At agent $l$, equally weighted particles $\big\{ \bd{x}_{k,n-1}^{\prime (j)} \big\}_{j=1}^{J}$ representing
the beliefs $b^{(P)}(\bd{x}_{k,n-1})$ with $k \rmv\in\rmv \{l\} \cup \OO$ are available (these were calculated at time $n-1$). At time $n$, 
agent $l$ performs the following operations. 

\vspace{1.5mm}

\emph{Step 1---Prediction}:\, From $\big\{ \bd{x}_{k,n-1}^{\prime (j)} \big\}_{j=1}^{J}$, PRs $\big\{ \bd{x}_{k,n}^{(j)} \big\}_{j=1}^{J}$ of the 
prediction messages $\phi_{\rightarrow n}(\bd{x}_{k,n})$, $k \rmv\in\rmv \{l\} \cup \OO$ are calculated via message filtering (see Section \ref{sec:messfilter}) 
based on the state-transition pdf $f(\bd{x}_{k,n}|\bd{x}_{k,n-1})$, i.e., for each $\bd{x}_{k,n-1}^{\prime (j)}$ one particle $\bd{x}_{k,n}^{(j)}$ 
is drawn from $f(\bd{x}_{k,n}|\bd{x}_{k,n-1}^{\prime (j)})$.

\vspace{2mm}

\emph{Step 2---BP message passing}:\, For each $k \rmv\in\rmv \{l\} \cup \OO$, the belief is initialized as 
$b^{(0)}(\bd{x}_{k,n}) =  \phi_{\rightarrow n}(\bd{x}_{k,n})$, in the sense that the PR of $\phi_{\rightarrow n}(\bd{x}_{k,n})$ is used as 
PR of $b^{(0)}(\bd{x}_{k,n})$. Then, for 
\vspace{.5mm}
$p = 1,\dots,P$:

\begin{enumerate}

\item[a)] For each $m \rmv\in\rmv \OO$, a PR $\big\{ \big( \bd{x}_{m,n}^{(j)},w_{m,n}^{(j)} \big) \big\}_{j=1}^{J}$ 
of $b^{(p)}(\bd{x}_{m,n})$ in \eqref{eq:marg_CoSLAT_T} is obtained via importance sampling 
\pagebreak 
with proposal distribution 
$\phi_{\rightarrow n}(\bd{x}_{m,n})$ (see Section \ref{sec:calc_belief_m}). That is, using the particles $\big\{ \bd{x}^{(j)}_{m,n} \big\}_{j=1}^{J}$ 
representing $\phi_{\rightarrow n}(\bd{x}_{m,n})$ (calculated in Step 1), nonnormalized weights are calculated as 
$\tilde{w}_{m,n}^{(j)} \rmv= \Phi_{m,n}^{(p)}(\bd{x}^{(j)}_{m,n})$ 
(cf.\ \eqref{eq:weight_m}) for all $j \in \{1,\dots,J\}$ in a distributed manner as described in Section \ref{sec:calcobjectmarg}. 
The final weights $w_{m,n}^{(j)}$ are obtained by a normalization.

\vspace{1mm}

\item[b)] (Not done for $p \!=\! P$) For each $m \!\in\! \cl{M}^{\OO}_{l,n}$, a PR of $\psi^{(p)}_{m\rightarrow l}(\bd{x}_{m,n})$ is 
calculated in a similar manner (see Section \ref{sec:extrinfo}).

\vspace{1mm}

\item[c)] A PR $\big\{ \big( \bd{x}_{l,n}^{(j)},w_{l,n}^{(j)} \big) \big\}_{j=1}^{J}$ of $b^{(p)}(\bd{x}_{l,n})$ 
is calculated by implementing \eqref{eq:marg_CoSLAT_agent} as described in Section \ref{sec:agent_beliefs}.
This involves equally weighted particles of all $b^{(p-1)}(\bd{x}_{l'\!,n})$, $l' \!\in\rmv \cl{M}^{\AA}_{l,n}$ 
(which were received from agents $l' \!\in\rmv \cl{M}^{\AA}_{l,n}$ at message passing iteration $p \!-\! 1$)
and of all $\psi^{(p-1)}_{m\rightarrow l}(\bd{x}_{m,n})$, $m \rmv\in\rmv \cl{M}^{\OO}_{l,n}$ (which were calculated
in Step 2b at message passing iteration $p \!-\! 1$).

\vspace{1mm}

\item[d)] 
 (Not done for $p \!=\! P$) For each $m \rmv\in\rmv \cl{M}^{\OO}_{l,n}$, a PR of $\psi^{(p)}_{l\rightarrow m}(\bd{x}_{l,n})$ 
is calculated in a similar manner.

\vspace{1mm}

\item[e)] For all PRs calculated in Steps 2a--2d, resampling is performed to obtain equally weighted particles.

\vspace{1mm}

\item[f)] (Not done for $p \!=\! P$) The equally weighted particles of $b^{(p)}(\bd{x}_{l,n})$ calculated in Step 2e are broadcast to all agents $l'$ 
for which $l \rmv\in\rmv \cl{M}^{\AA}_{l'\!,n}$, and equally weighted particles of $b^{(p)}(\bd{x}_{l_1\!,n})$ are received from each neighboring agent 
$l_1 \!\in\rmv \cl{M}^{\AA}_{l,n}$. Thus, at this point, agent $l$ has available equally weighted particles $\big\{ \bd{x}_{k,n}^{\prime (j)} \big\}_{j=1}^{J}$ of 
$b^{(p)}(\bd{x}_{k,n})$, $k \rmv\in \{l\} \cup \OO \cup \cl{M}_{l,n}^\AA$ and equally weighted particles $\big\{ \bd{x}_{m,n}^{\prime (j)} \big\}_{j=1}^{J}$ of
$\psi^{(p)}_{m \rightarrow l}(\bd{x}_{m,n})$ and $\big\{ \bd{x}_{l,n}^{\prime (j)} \big\}_{j=1}^{J}$ of 
$\psi^{(p)}_{l \rightarrow m}(\bd{x}_{l,n})$, $m \rmv\in\rmv \cl{M}^{\OO}_{l,n}$.

\end{enumerate}

\vspace{1.4mm}

\emph{Step 3---Estimation}:\, 
For $k \rmv\in\rmv \{l\} \rmv\cup\rmv \OO$, an approximation of the global MMSE state estimate $\hat{\bd{x}}^\text{MMSE}_{k,n}$ 
in \eqref{eq:mmse_slat} is computed from the PR $\big\{ \big( \bd{x}_{k,n}^{(j)},w_{k,n}^{(j)} \big) \big\}_{j=1}^{J}$ of $b^{(P)}(\bd{x}_{k,n})$ according 
\vspace{-1.8mm}
to \eqref{eq:et_part}, i.e.,
\[
\hat{\bd{x}}_{k,n} = \sum_{j=1}^J w_{k,n}^{(j)} \bd{x}_{k,n}^{(j)} \,, \quad k \rmv\in\rmv \{l\} \rmv\cup\rmv \OO.
\vspace{-1mm}
\]
}

\hrule 
\vspace{2mm}

\section{Variations and Implementation Aspects}
\label{sec:varImp}

Next, we discuss some variations and implementation aspects of the proposed algorithm.

\vspace{-1mm}

\subsection{Local Distributed Tracking}
\label{sec:LDT}

The convergence of the consensus or gossip algorithms used to calculate \eqref{eq:Phi_consensus} is slow if $|\AA_{m,n}| \ll |\AA|$, because then many initial consensus states $\zeta_{l,m;n}^{(j,0)}$ in \eqref{eq:CoSlatInit} are zero. We therefore introduce a modification, termed \emph{local distributed tracking} (LDT), in which  $b^{(p)}(\mathbf{x}_{m,n})$ for an object  $m \rmv\in\rmv \mathcal{O}$ is calculated via \eqref{eq:marg_CoSLAT_T} only at agents $l$ that acquire a measurement of the object, i.e., $l \rmv\in\rmv \AA_{m,n}$. The convergence is here significantly faster due to the smaller ``consensus network'' ($l \rmv\in\rmv \AA_{m,n}$ instead of $l \rmv\in\rmv \AA$) and the fact that zero initial values are avoided. LDT presupposes that the communication graph of the network formed by all agents $l \in \AA_{m,n}$ is connected. To ensure that agents $l' \in \AA_{m,n+1} \backslash \AA_{m,n} $  (i.e., $l' \!\notin\rmv \AA_{m,n}$ but $l' \!\in\rmv \AA_{m,n+1}$) obtain the information needed to track object $m$ at time $n+\rmv 1$, each agent $l \rmv\in\rmv \AA_{m,n}$ broadcasts $b^{(P)}(\mathbf{x}_{m,n})$ (calculated as described in Section \ref{sec:object_belief}) to its neighbors $l' \!\rmv\in\rmv \mathcal{C}_{l,n}$.
 
Using $b^{(P)}(\mathbf{x}_{m,n})$, neighboring agents $l' \!\in\rmv \AA_{m,n+1} \!\setminus\rmv \AA_{m,n}$ are then able to calculate $\phi_{\rightarrow n+1}(\mathbf{x}_{m,n+1})$ (see \eqref{eq:pred_mess} and Section \ref{sec:messfilter}) and to track object $m$ at time $n+\rmv 1$ according to \eqref{eq:marg_CoSLAT_T}.

LDT has certain drawbacks. First, only agents $l \rmv\in\rmv \AA_{m,n}$ obtain an estimate of the state of object $m$. (Equivalently, each agent $l \rmv\in\rmv \AA$ tracks only objects $m \rmv\in\rmv \mathcal{M}^{\mathcal{O}}_{l,n}$.) Second, the size of the consensus network, $|\AA_{m,n}|$, has to be estimated at each time $n$.
Third, in agent networks with few communication links, it is possible that an agent $l' \!\in\rmv \AA_{m,n+1} \!\setminus\rmv \AA_{m,n}$ cannot communicate with any agent $l \rmv\in\rmv \AA_{m,n}$ at time $n$, i.e., $l \!\notin\rmv \mathcal{C}_{l'\!,n}$. Then, agent $l'$ does not obtain $b^{(P)}(\mathbf{x}_{m,n})$ and cannot 
track object $m$ at time $n+ \rmv1$, even though it acquired a corresponding measurement at time $n$. However, in many scenarios, the communication regions of the agents are significantly larger than their measurement regions. The situation described above is then very unlikely.

\vspace{-.5mm}

\subsection{Alternative Proposal Distribution} 
\label{sec:CoSLAT_inital}

\vspace{.3mm}

In the proposed particle-based message multiplication scheme presented in Section \ref{sec:NBPlowcomplexity}, a PR of $b^{(p)}(\mathbf{x}_{k,n})$, $k \in \cl{E}$ is calculated via importance sampling using a proposal distribution $q(\bar{\mathbf{x}}_{k,n})$ (as given for $k=l \in \cl{A}$ by \eqref{eq:proposal}). However, this proposal distribution is not appropriate if the prediction message $\phi_{\rightarrow n}(\mathbf{x}_{k,n}) = \int f(\mathbf{x}_{k,n}|\mathbf{x}_{k,n-1}) \ist f(\mathbf{x}_{k,n-1}) \, d\mathbf{x}_{k,n-1}$ is not very informative. An uninformative proposal distribution implies that the generated particles will be widely spread and the estimation performance will be poor if only a moderate number of particles $J$ is used.

We therefore propose an alternative proposal distribution for entities $k \in \EE$ with an uninformative prediction message $\phi_{\rightarrow n}(\mathbf{x}_{k,n})$. This proposal distribution leads to accurate estimates even if a moderate number of particles $J$ is used. In the following, we will present the alternative proposal distribution for the calculation of an object belief $b^{(p)}(\bd{x}_{m,n})$, $m\in\OO$; a similar proposal distribution can be used for the calculation of an agent belief $b^{(p)}(\bd{x}_{l,n})$, $l\in\AA$. We start by recalling that in the original particle-based message multiplication scheme presented in Section \ref{sec:NBPlowcomplexity}, particle-based calculation of $b^{(p)}(\bd{x}_{m,n})$ relies on particle-based calculation of the ``stacked'' belief $b^{(p)}(\bar{\bd{x}}_{m,n})$, and this stacked belief is obtained by using \eqref{eq:meas_mess_3} in \eqref{eq:marg_CoSLAT_T} similarly as $b^{(p)}(\bar{\bd{x}}_{l,n})$, $l \in \cl{A}$ is obtained in Section \ref{sec:NBPlowcomplexity}. Furthermore, in analogy to \eqref{eq:proposal}, the proposal distribution for particle-based calculation of $b^{(p)}(\bar{\bd{x}}_{m,n})$ reads $q(\bar{\bd{x}}_{m,n}) \triangleq \phi_{\rightarrow n}(\bd{x}_{m,n}) \prod_{l \in \cl{A}_{m,n}} \psi^{(p-1)}_{l \rightarrow m}(\bd{x}_{l,n})$.

As an alternative to $q(\bar{\bd{x}}_{m,n})$, we now use the proposal distribution
\begin{equation}
\label{eq:q_mod}
\tilde{q}(\bar{\bd{x}}_{m,n}) \,\triangleq\, \phi^{(p-1)}_{\hat{l} \rightarrow m}(\bd{x}_{m,n}) \rrmv\rmv \prod_{l \in \cl{A}_{m,n} \backslash \{\hat{l}\}} \!\!\rmv\rmv \psi^{(p-1)}_{l \rightarrow m}(\bd{x}_{l,n}) \,,
\vspace{-.7mm}
\end{equation}
with some judiciously chosen $\hat{l} \!\in\rmv \AA_{m,n}$ (the precise choice of $\hat{l}$ will be discussed later). Based on \eqref{eq:meas_mess_3}, equally weighted particles representing $\phi^{(p)}_{\hat{l} \rightarrow m}(\mathbf{x}_{m,n})$ can be calculated from equally weighted particles representing $\psi_{\hat{l} \rightarrow m}^{(p-1)}(\bd{x}_{\hat{l},n})$ by a variant of the message filtering procedure described in Section \ref{sec:messfilter} (see also \cite[Section IV-C]{ihler}).

A PR $\big\{ \big(\bar{\mathbf{x}}_{m,n}^{(j)} , w_{m,n}^{(j)} \big) \big\}_{j=1}^{J}$ of $b^{(p)}(\bar{\mathbf{x}}_{m,n})$, $m \rmv\in\rmv \mathcal{O}$ is now obtained by means of importance sampling using the proposal distribution $\tilde{q}(\bar{\mathbf{x}}_{m,n})$ in \eqref{eq:q_mod}: 
for particles $\big\{ \bar{\mathbf{x}}_{m,n}^{(j)} \big\}_{j=1}^{J}$ drawn from $\tilde{q}(\bar{\mathbf{x}}_{m,n})$, weights $\big\{ w_{m,n}^{(j)} \big\}_{j=1}^{J}$ are calculated by evaluating $\tilde{w}_{m,n}^{(j)} \propto b^{(p)}(\bar{\bd{x}}^{(j)}_{m,n})/\tilde{q}(\bar{\bd{x}}^{(j)}_{m,n})$ followed by a normalization. 
Inserting \eqref{eq:marg_CoSLAT_T} and \eqref{eq:meas_mess_3} for $b^{(p)}(\bar{\bd{x}}^{(j)}_{m,n})$, and \eqref{eq:q_mod} for $\tilde{q}(\bar{\bd{x}}^{(j)}_{m,n})$, the nonnormalized weights are obtained as
\[
\tilde{w}^{(j)}_{m,n} = \frac{\phi_{\rightarrow n}(\mathbf{x}^{(j)}_{m,n})\prod_{l \in \cl{A}_{m,n} \backslash \{\hat{l}\}} f(\bd{y}_{l,m;n}|\bd{x}^{(j)}_{l,n}, \bd{x}^{(j)}_{m,n})}{\phi^{(p-1)}_{\hat{l} \rightarrow m}(\bd{x}^{(j)}_{m,n})}.
\]
Here, the messages $\phi_{\rightarrow n}(\mathbf{x}_{m,n})$ and $\phi^{(p-1)}_{\hat{l} \rightarrow m}(\bd{x}_{m,n})$ are evaluated by means of a Monte Carlo approximation \cite{doucet} of the integral in \eqref{eq:pred_mess} and \eqref{eq:meas_mess_3}, respectively (cf. \eqref{eq:phi_approx}).

To make $\tilde{q}(\bar{\mathbf{x}}_{m,n})$ in \eqref{eq:q_mod} maximally informative, we choose 
\begin{equation}
\label{eq:propindex}
\hat{l} =\, \underset{l \in \AA_{m,n}}{\mathrm{argmin}} \; \sigma^2_{l} \,,
\vspace{-1mm}
\end{equation}
where $\sigma^2_{l}$ is the empirical variance of $b^{(p)}_{l}(\mathbf{x}_{l,n})$.
For non-anchor agents, $\sigma^2_{l}$ is calculated from 
equally weighted particles $\big\{ \mathbf{x}_{l,n}^{(j)} \big\}_{j=1}^{J}$ representing $b^{(p)}_{l}(\mathbf{x}_{l,n})$ as
$\sigma^2_{l} = \frac{1}{J}\sum_{j = 1}^{J} \big\| \mathbf{x}^{(j)}_{l,n} \!-\rmv \bm{\mu}_{l} \big\|^2\rmv\rmv$,
with $\bm{\mu}_{l} = \frac{1}{J}\sum_{j = 1 }^{J} \rmv\rmv \mathbf{x}^{(j)}_{l,n}$. For anchors, we set $\sigma^2_{l} = 0$.

For a distributed implementation, \eqref{eq:propindex} is computed using the min-consensus scheme \cite{olfatisaber}, 
which converges in $I$ iterations \cite{olfatisaber} (in the case of LDT, the number of iterations is not $I$ but the diameter of the agent network given by $\AA_{m,n}$).
This min-consensus scheme is also used to disseminate the particles representing the optimum $\phi^{(p)}_{\hat{l} \rightarrow m}(\mathbf{x}_{m,n})$ within the network.

\section{Communication Requirements and Delay}
\label{sec:CoSLAT_comm}

\vspace{.5mm}

In the following discussion of the communication requirements of the proposed distributed CS--DT algorithm, we assume for simplicity
that all $\tilde{\bd{x}}_{k,n}$, $k\rmv\in\rmv \EE$ (i.e., the substates actually involved in the measurements, cf.\  \eqref{eq:mess})
have identical dimension $L$. Furthermore, we denote by $C$ the number of consensus or gossip iterations used for averaging,
by $P$ the number of message passing iterations, by $J$ the number of particles, and by $I$ the diameter of the communication graph.
For an analysis of the delay caused by communication, we assume that all agents can transmit in parallel. More specifically, 
broadcasting the beliefs of all the agents will be counted as one delay time slot, and broadcasting all quantities related to
one consensus iteration for averaging or maximization will also be counted as one delay time slot.

\begin{itemize}

\vspace{.5mm}

\item For calculation of the object beliefs $b^{(p)}(\bd{x}_{m,n})$, $m \rmv\in\rmv \OO$ using the ``consensus--over--weights'' scheme 
(see Section \ref{sec:object_belief} and Step 2a in Algorithm 1), at each time $n$, agent $l \rmv\in\rmv \AA$ broadcasts 
$N^{\text{C}} \rmv\triangleq P(C+I)J|\OO|$ real values to agents $l' \!\rmv\in\rmv \cl{C}_{l,n}$. The corresponding contribution to the overall delay
is $P(C+I)$ time slots, because the consensus coefficients for all objects are broadcast in parallel.

\vspace{1mm}

\item To support neighboring agents $l'$ with $l\rmv\in \cl{M}^{\AA}_{l'\!,n}$ in calculating their own beliefs $b^{(p)}(\bd{x}_{l'\!,n})$ 
(see Section \ref{sec:agent_beliefs} and Step 2c in Algorithm 1), at each time $n$, agent $l \rmv\in\rmv \AA$ broadcasts 
$N^{\text{NBP}} \rmv\triangleq\rmv PJL$ real values to those neighboring agents (see Step 2f in Algorithm 1). The delay contribution 
is $P$ time slots, 
because each agent broadcasts a belief in each message passing iteration.

\vspace{1mm}

\item At those times $n$ where the alternative proposal distribution described in Section \ref{sec:CoSLAT_inital} is used, agent $l \rmv\in\rmv \AA$ broadcasts
$N^{\text{AP}} \!\triangleq P J L I |\OO|$ real values to each agent $l' \!\rmv\in\rmv \mathcal{C}_{l,n}$ (in addition to $N^{\text{C}}$ and $N^{\text{NBP}}$). For disseminating the proposal distribution of each object, $I$ consensus iterations at each of the $P$ message passing iterations are needed. This results in a contribution to the overall delay of $PI$ time slots.

\vspace{1mm}

\end{itemize}

Therefore, at each time $n$, the total number of real values broadcast by each agent during $P$ message passing iterations is
\vspace{-.8mm}
\[
N^{\text{TOT}} \rmv= N^{\text{C}} \rmv+ N^{\text{NBP}} \rmv= PJ \ist \big((C \rmv+\rmv I) \ist |\OO| + L \big)\ist.
\]
The corresponding delay is $P(C+I)$ time slots. However, if the alternative proposal distribution is used,
\[
N^{\text{TOT}} \rmv= N^{\text{NBP}} \rmv+\rmv N^{\text{C}} \rmv+\rmv N^{\text{AP}} \rmv= PJ \ist \big( (C + I + L I)|\OO| + L \big) \ist,
\]
and the corresponding delay is
$P(C+I) + PI = P(C+2I)$ time slots.

\vspace{-1mm}

\section{Simulation Results}
\label{sec:simres}

We will study the performance and communication requirements of the proposed method (PM) in two dynamic scenarios and a static scenario. 
In addition, we will investigate the scalability of the PM by comparing it with two different particle filters and with conventional
nonparametric BP \cite{ihler,lien}. Simulation source files and animated plots are available at http://www.nt.tuwien.ac.at/about-us/staff/florian-meyer/.

\vspace{-.5mm}

\subsection{Dynamic Scenarios}
\label{sec:dynScenario}

\vspace{.5mm}

\begin{figure}
\vspace{1mm}
\centering
\psfrag{s01}[t][t][0.9]{\color[rgb]{0,0,0}\setlength{\tabcolsep}{0pt}\begin{tabular}{c}\raisebox{-1.3mm}{$x_1$}\end{tabular}}
\psfrag{s02}[b][b][0.9]{\color[rgb]{0,0,0}\setlength{\tabcolsep}{0pt}\begin{tabular}{c}\raisebox{.1mm}{$x_2$}\end{tabular}}
\psfrag{x01}[t][t][0.7]{$0$}
\psfrag{x02}[t][t][0.7]{$10$}
\psfrag{x03}[t][t][0.7]{$20$}
\psfrag{x04}[t][t][0.7]{$30$}
\psfrag{x05}[t][t][0.7]{$40$}
\psfrag{x06}[t][t][0.7]{$50$}
\psfrag{x07}[t][t][0.7]{$60$}
\psfrag{x08}[t][t][0.7]{$70$}
\psfrag{v01}[r][r][0.7]{$0$}
\psfrag{v02}[r][r][0.7]{$10$}
\psfrag{v03}[r][r][0.7]{$20$}
\psfrag{v04}[r][r][0.7]{$30$}
\psfrag{v05}[r][r][0.7]{$40$}
\psfrag{v06}[r][r][0.7]{$50$}
\psfrag{v07}[r][r][0.7]{$60$}
\psfrag{v08}[r][r][0.7]{$70$}\
\includegraphics[scale=0.41]{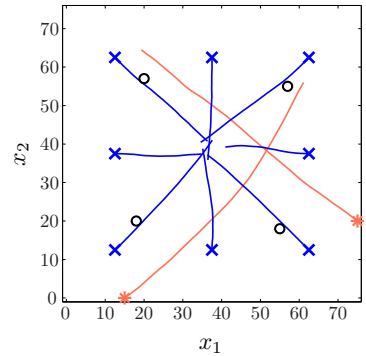}
\vspace{1mm}
\renewcommand{\baselinestretch}{1}\small\normalsize
\caption{Topology in the dynamic scenarios, with example trajectories of mobile agents and objects. 
Initial mobile agent locations are indicated by crosses, initial object locations by stars, and anchor locations by circles.}
\label{fig:top}
\vspace{-1.5mm}
\end{figure}

In the dynamic scenarios, we consider $|\AA|\!=\!12$ agents and $|\OO|\!=\!2$ objects as shown in Fig.\ \ref{fig:top}. Eight agents are mobile 
and four are static anchors (i.e., agents with perfect location information). Each agent has a communication range of 50 and attempts to localize itself 
(except for the anchors) and the two objects. The states of the mobile agents (MAs) and objects consist of location and velocity, i.e., 
$\bd{x}_{k,n} \!\triangleq\rmv ( x_{1,k,n}\,\,\ist x_{2,k,n}$\linebreak  
$\dot{x}_{1,k,n} \,\,\ist \dot{x}_{2,k,n} )^\text{T}\rmv$. All agents $l \in \AA$ acquire distance measurements according to \eqref{eq:mess}, i.e., 
$y_{l,k;n} = \|\tilde{\bd{x}}_{l,n} \!-\rmv \tilde{\bd{x}}_{k,n} \| + v_{l,k;n}\hspace{.15mm}$, where 
$\tilde{\bd{x}}_{k,n} \!\triangleq\rmv ( x_{1,k,n}\,\,\ist x_{2,k,n} )^{\text{T}}$ is the location of agent or object $k\!\in\! \cl{M}_{l,n}$ and the 
measurement noise $v_{l,k;n}$ is independent across $l$, $k$, and $n$ and Gaussian with variance $\sigma_v^2 \!=\! 2$. 
The states of the MAs and objects evolve independently according to $\bd{x}_{k,n} = \bd{G}\bd{x}_{k,n-1} + \bd{W}\bd{u}_{k,n}$, $n \!=\! 1,2,\dots$ \cite{rong}, where 
\[
\bd{G} = 
{ \begin{pmatrix}
   1 & 0 & 1 & 0 \\
   0 & 1 & 0 & 1 \\
   0  & 0  & 1 & 0 \\
   0 & 0 & 0 & 1
  \end{pmatrix} } \ist,
\quad\, \bd{W} =
{ \begin{pmatrix}
   0.5 & 0 \\
   0 & 0.5 \\
   1  & 0  \\
   0 & 1 
  \end{pmatrix} }\ist.
\vspace{-.5mm}
\]
The driving noise vectors $\bd{u}_{k,n} \!\in\! \mathbb{R}^2\rmv$ are Gaussian, i.e., $\bd{u}_{k,n}$ 
$\sim\rmv \cl{N}(\bd{0},\sigma_u^2\bd{I})$, with component variance $\sigma_{u}^2 \!=\! 5 \rmv\cdot\rmv 10^{-5}$ for the MAs and 
$\sigma_{u}^2 \rmv=\rmv 5 \rmv\cdot\rmv 10^{-4}$ for the objects; furthermore, $\bd{u}_{k,n}$ and $\bd{u}_{k'\!,n'}$ are independent unless $(k,n) \!=\! (k'\!,n')$. 
Each MA starts moving only when it is sufficiently localized in the sense that the empirical component variance of the estimated location vector 
is below $5\ist\sigma_v^2 \rmv=\rmv 10$; it then attempts to reach the center of the scene, $\tilde{\bd{x}}_{\text{c}} \!=\! ( 37.5\;\,\ist 37.5 )^\text{T}\rmv$, 
in 75 time steps. For generating the measurements, the MA trajectories are initialized using a Dirac-shaped prior $f(\bd{x}_{l,0})$
located at $\bm{\mu}_{l,0} \!=\! \big( x_{1,l,0} \,\,\, x_{2,l,0} \,\,\, (\tilde{x}_{1,\text{c}} \!-\! x_{1,l,0})/75 \,\,\, (\tilde{x}_{2,\text{c}} \!-\! x_{2,l,0})/75 \big)^\text{T}\rmv$, 
where $x_{1,l,0}$ and $x_{2,l,0}$ are chosen as shown in Fig.\ \ref{fig:top}, and the two object trajectories are initialized using a Dirac-shaped prior $f(\bd{x}_{m,0})$
located at $( 15 \;\, 0 \;\, 0.8 \;\, 0.6)^\text{T}\rmv$ and $(75 \,\,\ist 20 \,\,\ist {-0.8} \,\,\ist 0.6)^\text{T}\rmv$ (see Fig.\ \ref{fig:top}). Note that knowledge of 
these initial locations and velocities is not used by the simulated algorithms.

We compare the PM with a reference method (RM) that separately performs CS by means of a particle-based implementation of \cite{wymeersch} 
using nonparametric BP \cite{lien, ihler} and DT by means of a consensus-based distributed particle filter \cite{farahmand11, savic14}; 
the latter uses the MA location estimates provided by CS. Both PM and RM use $P \!=\! 1$ message passing iteration and $J \!=\! 1000$ particles. 
(We chose $P \!=\! 1$ since we observed almost no performance gains for $P > 1$; moreover, $P > 1$ may lead to overconfident beliefs \cite{froehle15}.)
For a distributed implementation, PM and RM employ $C \!=\! 6$ iterations of an average consensus with Metropolis weights \cite{xiao03}. 
They are initialized with a location prior for objects and MAs that is uniform on $[-200,200] \!\times\! [-200,200]$ 
and a Gaussian velocity prior for the MAs (after the respective MA is sufficiently localized as described above) with mean
 $\big( (\tilde{x}_{1,\text{c}} \!-\! \hat{\tilde{x}}_{1,l,n'})/75 \,\,\, (\tilde{x}_{2,\text{c}} \!-\! \hat{\tilde{x}}_{2,l,n'})/75 \big)^\text{T}\rmv$ 
and covariance matrix $\mathrm{diag}\ist\{10^{-3}\rmv, 10^{-3}\}$. Here, $\hat{\tilde{\bd{x}}}_{l,n'}$ is the location estimate at the time $n'$ 
at which MA $l$ is sufficiently localized for the first time. We furthermore used a velocity prior for the objects that is Gaussian with mean $\bm{\mu}_{m,0}^{(\text{v})}$ 
and covariance $\bd{C}_{m,0}^{(\text{v})}$. Here, $\bd{C}_{m,0}^{(\text{v})} \rmv=\rmv \mathrm{diag}\ist\{0.001, 0.001\}$ represents the uncertainty 
in knowing the velocity $\dot{\tilde{\bd{x}}}_{m,0}$ of object $m$ at time $n = 0$, and $\bm{\mu}_{m,0}^{(\text{v})}$ is a random hyperparameter 
that was sampled for each simulation run from $\cl{N}(\dot{\tilde{\bd{x}}}_{m,0},\bd{C}_{m,0}^{(\text{v})})$.

We simulated two different dynamic scenarios. In dynamic scenario 1, the measurement range of those four MAs that are initially located near the corners 
as shown in Fig.\ \ref{fig:top} (these agents will be termed ``corner agents'') is limited as specified later whereas all the other agents cover the entire field 
of size $75\ist\times\ist 75$. In dynamic scenario 2, the measurement range of all agents is limited to $20$. In both dynamic scenarios, for particle-based 
message multiplication (cf.\ Section \ref{sec:NBPlowcomplexity}) at time $n=1$, we used the alternative proposal distribution described in Section \ref{sec:CoSLAT_inital}.
We note that these dynamic scenarios cannot be tackled by SLAT algorithms \cite{taylor,funiak,savic15, kantas12, teng2012distr, uney14} since they involve MAs 
whereas SLAT assumes static agents.

\begin{figure}[t]
\centering
\psfrag{s01}[t][t][0.8]{\color[rgb]{0,0,0}\setlength{\tabcolsep}{0pt}\begin{tabular}{c}\raisebox{-1.3mm}{$n$}\end{tabular}}
\psfrag{s02}[b][b][0.8]{\color[rgb]{0,0,0}\setlength{\tabcolsep}{0pt}\begin{tabular}{c}\vspace{-2.5mm}{RMSE}\end{tabular}}
\psfrag{s05}[l][l]{\color[rgb]{0,0,0}}
\psfrag{s06}[l][l][0.72]{\color[rgb]{0,0,0}MA self-localization RMSE of CS}
\psfrag{s07}[l][l][0.72]{\color[rgb]{0,0,0}\raisebox{0.5mm}{MA self-localization RMSE of PM}}
\psfrag{s08}[l][l][0.72]{\color[rgb]{0,0,0}Object localization RMSE of RM}
\psfrag{s09}[l][l][0.72]{\color[rgb]{0,0,0}Object localization RMSE of PM}
\psfrag{s11}[][]{\color[rgb]{0,0,0}\setlength{\tabcolsep}{0pt}\begin{tabular}{c} \end{tabular}}
\psfrag{s12}[][]{\color[rgb]{0,0,0}\setlength{\tabcolsep}{0pt}\begin{tabular}{c} \end{tabular}}
\psfrag{x01}[t][t][0.72]{$1$}
\psfrag{x02}[t][t][0.72]{$10$}
\psfrag{x03}[t][t][0.72]{$20$}
\psfrag{x04}[t][t][0.72]{$30$}
\psfrag{x05}[t][t][0.72]{$40$}
\psfrag{x06}[t][t][0.72]{$50$}
\psfrag{x07}[t][t][0.72]{$60$}
\psfrag{x08}[t][t][0.72]{$70$}
\psfrag{v01}[r][r][0.72]{$0\!$}
\psfrag{v02}[r][r][0.72]{$2$}
\psfrag{v03}[r][r][0.72]{$4$}
\psfrag{v04}[r][r][0.72]{$6$}
\psfrag{v05}[r][r][0.72]{$8$}
\psfrag{v06}[r][r][0.72]{$10$}
\psfrag{v07}[r][r][0.72]{$12$}
\includegraphics[height=47mm, width=78mm]{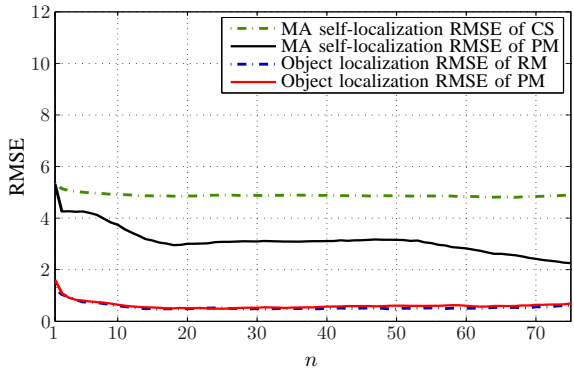}
\vspace{-.5mm}
\caption{MA self-localization RMSE and object localization RMSE versus time $n$ (dynamic scenario 1).}
\label{fig:plot}
\vspace{-.5mm}
\end{figure}

Fig.\ \ref{fig:plot} shows the root-mean-square errors (RMSEs) of MA self-localiza\-tion and object localization for $n \rmv=1,\ldots,75$ in dynamic scenario 1, 
with the measurement range of the corner agents chosen as $20$. The MA self-localiza\-tion RMSE and the object localization RMSE were determined 
by averaging over all MAs and all objects, respectively, and over 100 simulation runs. It is seen that the MA self-localization RMSE of 
PM is significantly smaller than that of RM. This is because with pure CS, the corner agents do not have enough partners for accurate self-localization, whereas with 
PM, they can use their measured distances to the objects to calculate the messages from the object nodes, $\phi^{(p)}_{m \rightarrow l}(\bd{x}_{l,n})$, which support
self-localization. The object localization RMSEs of PM and RM are very similar at all times. This is because the objects are always measured by several well-localized agents.

\begin{figure}[t]
\centering
\psfrag{s01}[t][t][0.75]{\color[rgb]{0,0,0}\setlength{\tabcolsep}{0pt}\begin{tabular}{c}\raisebox{-1.5mm}{$\rho$}\end{tabular}}
\psfrag{s02}[b][b][0.8]{\color[rgb]{0,0,0}\setlength{\tabcolsep}{0pt}\begin{tabular}{c}\vspace{-2.5mm}{RMSE}\end{tabular}}
\psfrag{s05}[l][l][0.72]{\color[rgb]{0,0,0}}
\psfrag{s06}[l][l][0.72]{\color[rgb]{0,0,0}MA self-localization RMSE of CS}
\psfrag{s07}[l][l][0.72]{\color[rgb]{0,0,0}\raisebox{0.5mm}{MA self-localization RMSE of PM}}
\psfrag{s08}[l][l][0.72]{\color[rgb]{0,0,0}Object localization RMSE of RM}
\psfrag{s09}[l][l][0.72]{\color[rgb]{0,0,0}Object localization RMSE of PM}
\psfrag{s11}[][]{\color[rgb]{0,0,0}\setlength{\tabcolsep}{0pt}\begin{tabular}{c} \end{tabular}}
\psfrag{s12}[][]{\color[rgb]{0,0,0}\setlength{\tabcolsep}{0pt}\begin{tabular}{c} \end{tabular}}
\psfrag{x01}[t][t][0.72]{$10$}
\psfrag{x02}[t][t][0.72]{$12.5$}
\psfrag{x03}[t][t][0.72]{$15$}
\psfrag{x04}[t][t][0.72]{$17.5$}
\psfrag{x05}[t][t][0.72]{$20$}
\psfrag{x06}[t][t][0.72]{$22.5$}
\psfrag{x07}[t][t][0.72]{$25$}
\psfrag{x08}[t][t][0.72]{$27.5$}
\psfrag{x09}[t][t][0.72]{$30$}
\psfrag{v01}[r][r][0.72]{$0$}
\psfrag{v02}[r][r][0.72]{$2$}
\psfrag{v03}[r][r][0.72]{$4$}
\psfrag{v04}[r][r][0.72]{$6$}
\psfrag{v05}[r][r][0.72]{$8$}
\psfrag{v06}[r][r][0.72]{$10$}
\psfrag{v07}[r][r][0.72]{$12$}
\includegraphics[height=47mm, width=78mm]{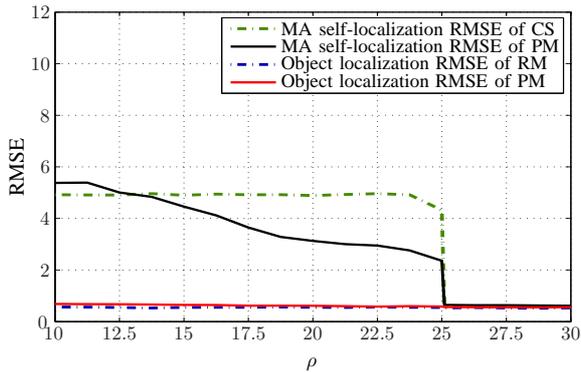}
\vspace{.7mm}
\caption{MA self-localization RMSE and object localization RMSE versus measurement range $\rho$ of the corner agents (dynamic scenario 1).}
\label{fig:plot_var}
\vspace{-1mm}
\end{figure}

\begin{figure}[t]
\centering
\psfrag{s01}[t][t][0.8]{\color[rgb]{0,0,0}\setlength{\tabcolsep}{0pt}\begin{tabular}{c}\raisebox{-1.3mm}{$n$}\end{tabular}}
\psfrag{s02}[b][b][0.8]{\color[rgb]{0,0,0}\setlength{\tabcolsep}{0pt}\begin{tabular}{c}\vspace{-2.5mm}{RMSE}\end{tabular}}
\psfrag{s05}[l][l]{\color[rgb]{0,0,0}}
\psfrag{s06}[l][l][0.72]{\color[rgb]{0,0,0}MA self-localization RMSE of CS}
\psfrag{s07}[l][l][0.72]{\color[rgb]{0,0,0}\raisebox{0.5mm}{MA self-localization RMSE of PM}}
\psfrag{s08}[l][l][0.72]{\color[rgb]{0,0,0}Object localization RMSE of RM}
\psfrag{s09}[l][l][0.72]{\color[rgb]{0,0,0}Object localization RMSE of PM}
\psfrag{s11}[][]{\color[rgb]{0,0,0}\setlength{\tabcolsep}{0pt}\begin{tabular}{c} \end{tabular}}
\psfrag{s12}[][]{\color[rgb]{0,0,0}\setlength{\tabcolsep}{0pt}\begin{tabular}{c} \end{tabular}}
\psfrag{x01}[t][t][0.72]{$1$}
\psfrag{x02}[t][t][0.72]{$10$}
\psfrag{x03}[t][t][0.72]{$20$}
\psfrag{x04}[t][t][0.72]{$30$}
\psfrag{x05}[t][t][0.72]{$40$}
\psfrag{x06}[t][t][0.72]{$50$}
\psfrag{x07}[t][t][0.72]{$60$}
\psfrag{x08}[t][t][0.72]{$70$}
\psfrag{v01}[r][r][0.72]{$0\!$}
\psfrag{v02}[r][r][0.72]{$5$}
\psfrag{v03}[r][r][0.72]{$10$}
\psfrag{v04}[r][r][0.72]{$15$}
\psfrag{v05}[r][r][0.72]{$20$}
\psfrag{v06}[r][r][0.72]{$25$}
\psfrag{v07}[r][r][0.72]{$30$}
\includegraphics[height=47mm, width=78mm]{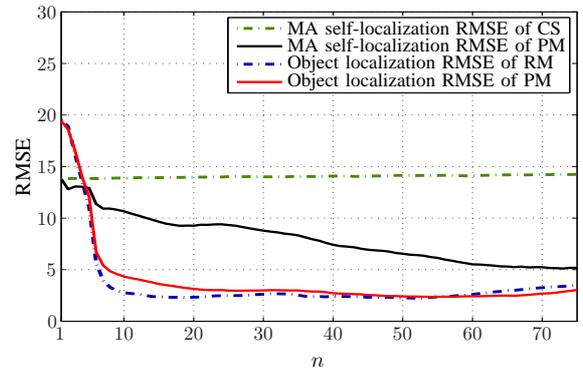}
\vspace{-.5mm}
\caption{MA self-localization RMSE and object localization RMSE versus time $n$ (dynamic scenario 2).}
\label{fig:plot2}
\vspace{-1.5mm}
\end{figure}

Still in dynamic scenario 1, Fig.\ \ref{fig:plot_var} shows the MA self-localiza\-tion and object localization RMSEs averaged over time $n$ versus the measurement range 
$\rho$ of the corner agents. For small and large $\rho$, PM performs similarly to RM but for different reasons: When $\rho$ is smaller than 12.5, 
the objects appear in the measurement regions of the corner agents only with a very small probability. Thus, at most times, the messages 
$\phi^{(p)}_{m \rightarrow l}(\bd{x}_{l,n})$ from the object nodes cannot be calculated. For $\rho$ larger than 25, the corner agents 
measure three well-localized agents at time $n \rmv=\! 1$, and thus they are also able to localize themselves using pure CS. However,
for $\rho$ between 15 and 25, PM significantly outperforms RM (cf.\ our discussion of Fig.\ \ref{fig:plot}). The object localization RMSEs of PM and RM 
are very similar and almost independent of $\rho$. This is because for all $\rho$, the objects are again measured by several well-localized agents.

Finally, Fig.\ \ref{fig:plot2} shows the MA self-localization and object localization RMSEs for $n \rmv=\rmv 1,\ldots,75$ in dynamic scenario 2 
(i.e., the measurement range of all agents is $20$). It can be seen that with both methods, the objects are roughly localized after a few initial time steps. 
However, with RM, due to the limited measurement range, not even a single MA can be localized. With PM, once meaningful probabilistic information 
about the object locations is available, also the self-localization RMSE decreases and most of the MAs can be localized after some time. 
This is possible since the MAs obtain additional information related to the measured objects. Which MAs are localized how well and at what times 
depends on the object trajectories and varies between the simulation runs.

Summarizing the results displayed in Figs.\ \ref{fig:plot}--\ref{fig:plot2}, one can conclude that the MA self-localization performance of PM is generally 
much better than that of RM whereas the object localization performance is not improved. In Section \ref{sec:statScenario}, we will present a scenario
in which also object localization is improved. The quantities determining the communication requirements of PM according to Section \ref{sec:CoSLAT_comm}
are $N^{\text{C}} \!=\!18000$ and $N^{\text{NBP}} \!=2000$. The resulting total communication requirement per MA and time step is 
$N^{\text{TOT}} = 20000$ for both dynamic scenarios. At time $n=1$, each MA additionally broadcasts $N^{\text{AP}} \rmv= 12000$ real values for disseminating the alternative proposal distribution. According to Section \ref{sec:CoSLAT_comm}, the delay for $n \geq 2$ is $9$ time slots in both dynamic scenarios; for $n=1$, the delay is $12$ time slots, due to the use of the alternative proposal distribution. RM has the same communication requirements and causes the same delay as 
PM.

\vspace{-.5mm}

\subsection{Static Scenario}
\label{sec:statScenario}

\vspace{.5mm}

\begin{figure}
\centering
\psfrag{s01}[t][t][0.9]{\color[rgb]{0,0,0}\setlength{\tabcolsep}{0pt}\begin{tabular}{c}\raisebox{-1.3mm}{$x_1$}\end{tabular}}
\psfrag{s02}[b][b][0.9]{\color[rgb]{0,0,0}\setlength{\tabcolsep}{0pt}\begin{tabular}{c}\raisebox{.1mm}{$x_2$}\end{tabular}}
\psfrag{x01}[t][t][0.7]{$0$}
\psfrag{x02}[t][t][0.7]{$20$}
\psfrag{x03}[t][t][0.7]{$40$}
\psfrag{x04}[t][t][0.7]{$60$}
\psfrag{x05}[t][t][0.7]{$80$}
\psfrag{x06}[t][t][0.7]{$100$}
\psfrag{v01}[r][r][0.7]{$0$}
\psfrag{v02}[r][r][0.7]{$20$}
\psfrag{v03}[r][r][0.7]{$40$}
\psfrag{v04}[r][r][0.7]{$60$}
\psfrag{v05}[r][r][0.7]{$80$}
\psfrag{v06}[r][r][0.7]{$100$}
\includegraphics[scale=0.41]{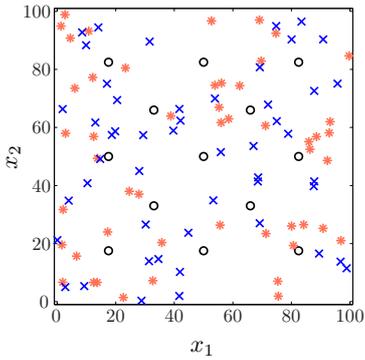}
\vspace{1mm}
\renewcommand{\baselinestretch}{1}\small\normalsize
\caption{Topology in the static scenario, with anchor locations (indicated by circles)
and example realizations of non-anchor agent locations (indicated by crosses) and of object locations (indicated by stars).}
\label{fig:topStatic}
\vspace{-1mm}
\end{figure}

Next, we consider a completely static scenario. This can be more challenging than the dynamic scenario considered in the last section, 
since at the first message passing iterations the beliefs of all entities can be highly multimodal, and thus BP algorithms using Gaussian 
approximations \cite{sathyan13, meyer14sigma} are typically not suitable for reliable localization. In the simulated scenario, there are
$|\AA|\!=\!63$ static agents and $|\OO|\!=\!50$ static objects. 13 agents are anchors located as depicted in Fig.\ \ref{fig:topStatic}. 
The 50 remaining agents and the objects are randomly (uniformly) placed in a field of size 100$\ist\times\ist$100; a realization of the locations 
of the non-anchor agents and objects is shown in Fig.\ \ref{fig:topStatic}. The states of the non-anchor agents and of the objects are the locations, i.e., 
$\bd{x}_{k,n} \!=\rmv \tilde{\bd{x}}_{k,n} \!=\rmv ( x_{1,k,n}\,\,\ist x_{2,k,n} )^{\text{T}}\rmv$. Each agent performs distance measurements according to 
\eqref{eq:mess} with a measurement range of 22.5 and a noise variance of $\sigma_v^2 \!=\! 2$. The communication range of each agent is 50.
The prior for the non-anchor agents and for the objects is uniform on $[-200,200] \!\times\! [-200,200]$. 
Both PM and RM use $J \!=\! 1000$ particles and $C \!=\! 15$ average consensus iterations.
Since all agents and objects are static, we simulated only a single time step. This scenario is similar to that considered in \cite{wymeersch} for pure CS,
except that 50 of the agents used in \cite{wymeersch} are replaced by objects and also anchor nodes perform measurements. For message multiplication,
we used the alternative proposal distribution described in Section \ref{sec:CoSLAT_inital}.

\begin{figure}
\centering
\psfrag{s01}[t][t][0.8]{\color[rgb]{0,0,0}\setlength{\tabcolsep}{0pt}\begin{tabular}{c}\raisebox{-1.5mm}{$p$}\end{tabular}}
\psfrag{s02}[b][b][0.8]{\color[rgb]{0,0,0}\setlength{\tabcolsep}{0pt}\begin{tabular}{c}\vspace{-4.5mm}{RMSE}\end{tabular}}
\psfrag{s05}[l][l][0.72]{\color[rgb]{0,0,0}}
\psfrag{s06}[l][l][0.72]{\color[rgb]{0,0,0}Object localization RMSE of RM}
\psfrag{s07}[l][l][0.72]{\color[rgb]{0,0,0}Agent self-localization RMSE of RM}
\psfrag{s08}[l][l][0.72]{\color[rgb]{0,0,0}Object localization RMSE of PM}
\psfrag{s09}[l][l][0.72]{\color[rgb]{0,0,0}Agent self-localization RMSE of PM}
\psfrag{s11}[][]{\color[rgb]{0,0,0}\setlength{\tabcolsep}{0pt}\begin{tabular}{c} \end{tabular}}
\psfrag{s12}[][]{\color[rgb]{0,0,0}\setlength{\tabcolsep}{0pt}\begin{tabular}{c} \end{tabular}}
\psfrag{x01}[t][t][0.72]{$1$}
\psfrag{x02}[t][t][0.72]{$2$}
\psfrag{x03}[t][t][0.72]{$3$}
\psfrag{x04}[t][t][0.72]{$4$}
\psfrag{x05}[t][t][0.72]{$5$}
\psfrag{x06}[t][t][0.72]{$6$}
\psfrag{x07}[t][t][0.72]{$7$}
\psfrag{v01}[r][r][0.72]{$0$}
\psfrag{v02}[r][r][0.72]{$2$}
\psfrag{v03}[r][r][0.72]{$4$}
\psfrag{v04}[r][r][0.72]{$6$}
\psfrag{v05}[r][r][0.72]{$8$}
\includegraphics[height=47mm, width=78mm]{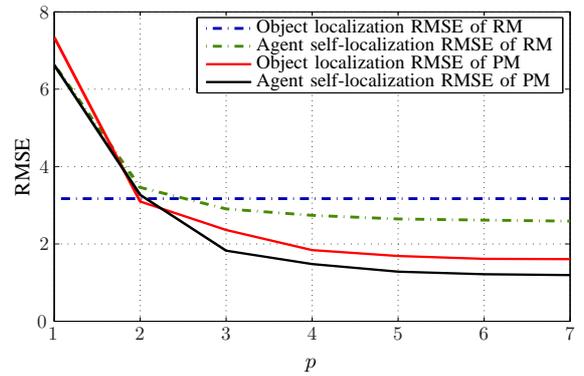}
\vspace{.5mm}
\caption{Non-anchor agent self-localization RMSE and object localization RMSE versus message passing iteration index $p$ (static scenario).}
\label{fig:plot_p}
\vspace{-.5mm}
\end{figure}

Fig.\ \ref{fig:plot_p} shows the localization RMSEs versus the message passing iteration index $p$. It is seen that the agent self-localization performance 
of PM is significantly better than that of RM. Again, this is because agents can use messages $\phi^{(p)}_{m \rightarrow l}(\bd{x}_{l,n})$ from well-localized 
objects to better localize themselves. Furthermore, also the object localization performance of PM is significantly better. This is because with separate 
CS and DT, poor self-localization of certain agents degrades the object localization performance. It is finally seen that increasing $p$ beyond 5 
does not lead to a significant reduction of the RMSEs.

In this scenario, assuming $P\rmv=\rmv3$, we have $N^{\text{C}} \rmv=\rmv 2.70\cdot 10^6$, $N^{\text{NBP}} \rmv=\rmv 6000$,
and $N^{\text{TOT}} \rmv=\rmv 2.71 \cdot 10^6$. For proposal adaptation, each non-anchor agent additionally broadcasts
$N^{\text{AP}} \rmv= 9.00 \cdot 10^5\rmv$ real values. The delay is $54$ time slots.

\vspace{-1mm}

\subsection{Scalability}
\label{sec:compParticle}

Finally, we consider again a dynamic scenario and investigate the scalability of PM for growing network size in comparison to conventional
particle filtering approximating the classical sequential Bayesian filter. Because this aspect is not fundamentally related to a distributed implementation,
we consider a centralized scenario where all measurements are processed at a fusion center. We compare a centralized version of PM with the 
sampling importance resampling particle filter \cite{arulampalam} (abbreviated as SPF), the unscented particle filter (UPF) \cite{merwe00}, and a 
particle implementation of the proposed BP scheme \eqref{eq:marg_CoSLAT_agent}--\eqref{eq:extr_infor_1} using conventional
nonparametric belief propagation (NBP) \cite{ihler,lien}. Because PM and NBP are centralized, they do not need a consensus for DT. Both SPF and 
UPF estimate the total ``stacked'' state of all MAs and objects, whose dimension grows with the network size. The state of an MA or object consists 
of location and velocity.

We consider mobile networks of increasing size $\big(|\tilde{\AA}|,|\OO|\big) = (8,2)$, $(16,4)$, $(32,8)$, $(64,16)$, and $(128,32)$, 
where $\tilde{\AA} \subseteq \AA$ is the set of MAs. In addition to the MAs and objects, four anchors are placed at locations $({-100} \; {-100})^\text{T}\rmv$, 
$({-100} \;\, 100)^\text{T}\rmv$, $(100 \;\, {-100})^\text{T}\rmv$, and $(100 \;\, 100)^\text{T}\rmv$.
For the MAs and objects, we use the motion model of Section \ref{sec:dynScenario} with driving noise variance $\sigma_u^2 \!=\! 10^{-2}\rmv$, 
and for the agents (MAs and anchors), we use the measurement model of Section \ref{sec:dynScenario} with measurement noise variance $\sigma_v^2 \!=\! 1$. 
For generating the measurements, the MA and object trajectories are initialized as 
$( x_{1,k,0} \,\,\, x_{2,k,0} \,\,\, 0 \,\,\, 0 )^\text{T}\rmv$, where $x_{1,k,0}$ and $x_{2,k,0}$ are randomly (uniformly) 
chosen in a field of size 100$\ist\times\ist$100. The algorithms are initialized with the initial prior pdf 
$f(\bd{x}_{k,0}) \rmv =\rmv \mathcal{N}(\bm{\mu}_{k,0},\bd{C}_{k,0} )$. Here, $\bd{C}_{k,0} =\!\mathrm{diag}\ist\{10^{-2}, 10^{-2},$ \linebreak 
$10^{-2},10^{-2}\}$, and $\bm{\mu}_{k,0}$ is sampled for each simulation run from $\mathcal{N}(\bd{x}_{k,0}^{\text{true}},\bd{C}_{k,0})$, 
where $\bd{x}_{k,0}^{\text{true}}$ is the true initial state used for generating an MA or object trajectory. Since an informative initial prior is available 
for all MAs and objects, we do not use the alternative proposal distribution.

The measurement topology of the network is randomly determined at each time step as follows: Each MA measures, with equal probability, 
one or two randomly chosen anchors and two randomly chosen MAs or objects. This is done such that each object is measured by two randomly chosen MAs. 
In addition, each object is also measured by one or two randomly chosen anchors. The sets $\cl{M}^{\tilde{\cl{A}}}_{l,n}$, $l \in \tilde{\cl{A}}$ are 
symmetric in that $l'\!\in\rmv \cl{M}^{\tilde{\cl{A}}}_{l,n}$ implies $l\rmv\in\rmv\cl{M}^{\tilde{\cl{A}}}_{l'\!,n}$. 
Furthermore, the sets $\cl{M}_{l,n}$, $l \in \tilde{\cl{A}}$ are chosen such that the topology graph that is constituted by the object measurements performed by the MAs 
corresponds to a Hamiltonian cycle, i.e., each MA and object in the graph is visited exactly once \cite{balakrishnan12}. 
This ensures that the expected number of neighbors of each MA and object is independent of the network size and the network is connected at all times.
PM, SPF, and UPF use $J \rmv=\rmv 1000$ and $5000$ particles, whereas NBP uses $J \rmv=\rmv 500$ and $1000$ particles
($J \rmv=\rmv 5000$ would have resulted in excessive runtimes, due to the quadratic growth of NBP's complexity with $J$).
PM and NBP use $P \rmv=\rmv 2$ message passing iterations. We performed 100 simulation runs, each consisting of 100 time steps.

\begin{figure}
\centering
\psfrag{s01}[t][t][0.8]{\color[rgb]{0,0,0}\setlength{\tabcolsep}{0pt}\begin{tabular}{c}\raisebox{-3mm}{numbers of agents and objects $(|\tilde{\cl{A}}|, |\cl{O}|)$}\end{tabular}}
\psfrag{s02}[b][b][0.8]{\color[rgb]{0,0,0}\setlength{\tabcolsep}{0pt}\begin{tabular}{c}\raisebox{0mm}{runtime (s)}\end{tabular}}
\psfrag{s05}[l][l][0.72]{\color[rgb]{0,0,0}}
\psfrag{s06}[l][l][0.72]{\color[rgb]{0,0,0}NBP 
($J \!=\! 1000$)}
\psfrag{s07}[l][l][0.72]{\color[rgb]{0,0,0}UPF 
($J \!=\rmv 5000$)}
\psfrag{s08}[l][l][0.72]{\color[rgb]{0,0,0}NBP 
($J \!=\rmv 500$)}
\psfrag{s09}[l][l][0.72]{\color[rgb]{0,0,0}UPF 
($J \!=\! 1000$)}
\psfrag{s10}[l][l][0.72]{\color[rgb]{0,0,0}PM 
($J \!=\rmv 5000$)}
\psfrag{s11}[l][l][0.72]{\color[rgb]{0,0,0}SPF 
($J \!=\rmv 5000$)}
\psfrag{s12}[l][l][0.72]{\color[rgb]{0,0,0}PM 
($J \!=\! 1000$)}
\psfrag{s13}[l][l][0.72]{\color[rgb]{0,0,0}SPF 
($J \!=\! 1000$)}
\psfrag{s15}[][]{\color[rgb]{0,0,0}\setlength{\tabcolsep}{0pt}\begin{tabular}{c} \end{tabular}}
\psfrag{s16}[][]{\color[rgb]{0,0,0}\setlength{\tabcolsep}{0pt}\begin{tabular}{c} \end{tabular}}

\psfrag{x01}[t][t][0.72]{$(8,2)$}
\psfrag{x02}[t][t][0.72]{$(16,4)$}
\psfrag{x03}[t][t][0.72]{$(32,8)$}
\psfrag{x04}[t][t][0.72]{$(64,16)$}
\psfrag{x05}[t][t][0.72]{$(128,32)$}

\psfrag{v01}[r][r][0.72]{$10^{0}$}
\psfrag{v02}[r][r][0.72]{$10^{1}$}
\psfrag{v03}[r][r][0.72]{$10^{2}$}
\psfrag{v04}[r][r][0.72]{$10^{3}$}
\psfrag{v05}[r][r][0.72]{$10^{4}$}
\includegraphics[height=47mm, width=78mm]{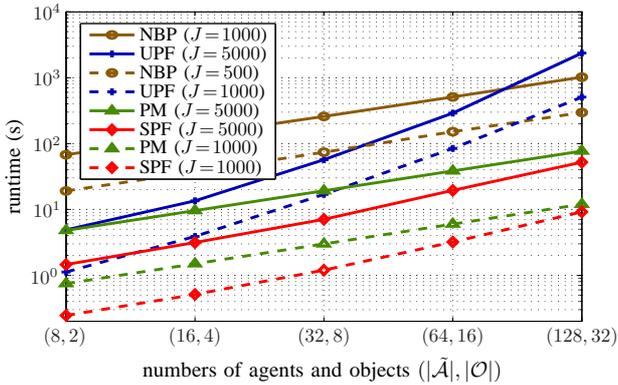}
\vspace{2mm}
\caption{Average runtime versus network size.}
\vspace{-2mm}
\label{fig:scalabilityRuntime}
\end{figure}

Fig.\ \ref{fig:scalabilityRuntime} shows the average runtime in seconds of all operations performed by the fusion center during one time step 
versus the network size $(|\tilde{\cl{A}}|, |\cl{O}|)$. The runtime was measured using MATLAB implementations of the algorithms on a single core 
of an Intel Xeon X5650 CPU. It is seen that the runtime of the BP-based methods PM and NBP scales linearly in the network size, whereas that 
of the particle filters SPF and UPF scales polynomially. This polynomial scaling of SPF and UPF is due to the fact that these filters perform 
operations involving matrices whose dimension increases with the network size. For the same number of particles $J$, PM always runs faster than UPF.  
Furthermore, for the considered parameters and network sizes, NBP has the highest runtime and SPF the lowest; however, for larger network sizes, 
the runtime of SPF and UPF will exceed that of PM and NBP due to the polynomial scaling characteristic of SPF and UPF. It is also seen that the 
runtime of PM is significantly lower than that of NBP.

\begin{figure}
\centering
\psfrag{s01}[t][t][0.8]{\color[rgb]{0,0,0}\setlength{\tabcolsep}{0pt}\begin{tabular}{c}\raisebox{-3mm}{numbers of agents and objects $(|\tilde{\cl{A}}|, |\cl{O}|)$}\end{tabular}}
\psfrag{s02}[b][b][0.8]{\color[rgb]{0,0,0}\setlength{\tabcolsep}{0pt}\begin{tabular}{c}RMSE\end{tabular}}
\psfrag{s05}[l][l][0.72]{\color[rgb]{0,0,0}}
\psfrag{s06}[l][l][0.72]{\color[rgb]{0,0,0}SPF 
($J \!=\! 1000$)}
\psfrag{s07}[l][l][0.72]{\color[rgb]{0,0,0}SPF 
($J \!=\rmv 5000$)}
\psfrag{s08}[l][l][0.72]{\color[rgb]{0,0,0}NBP 
($J \!=\rmv 500$)}
\psfrag{s09}[l][l][0.72]{\color[rgb]{0,0,0}NBP 
($J \!=\! 1000$)}
\psfrag{s10}[l][l][0.72]{\color[rgb]{0,0,0}PM 
($J \!=\! 1000$)}
\psfrag{s11}[l][l][0.72]{\color[rgb]{0,0,0}PM 
($J \!=\rmv 5000$)}
\psfrag{s12}[l][l][0.72]{\color[rgb]{0,0,0}UPF 
($J \!=\! 1000$)}
\psfrag{s13}[l][l][0.72]{\color[rgb]{0,0,0}UPF 
($J \!=\rmv 5000$)}
\psfrag{s15}[][]{\color[rgb]{0,0,0}\setlength{\tabcolsep}{0pt}\begin{tabular}{c} \end{tabular}}
\psfrag{s16}[][]{\color[rgb]{0,0,0}\setlength{\tabcolsep}{0pt}\begin{tabular}{c} \end{tabular}}

\psfrag{x01}[t][t][0.72]{$(8,2)$}
\psfrag{x02}[t][t][0.72]{$(16,4)$}
\psfrag{x03}[t][t][0.72]{$(32,8)$}
\psfrag{x04}[t][t][0.72]{$(64,16)$}
\psfrag{x05}[t][t][0.72]{$(128,32)$}

\psfrag{v01}[r][r][0.72]{$10^{0}$}
\psfrag{v02}[r][r][0.72]{$10^{1}$}
\includegraphics[height=47mm, width=78mm]{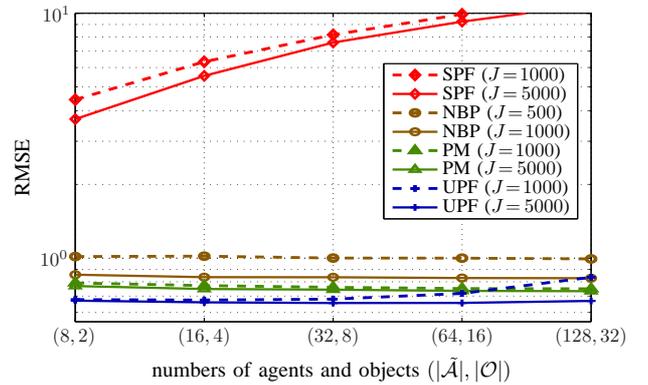}
\vspace{2.2mm}
\caption{Average localization RMSE versus network size.}
\vspace{-2.7mm}
\label{fig:scalabilityRMSE}
\end{figure}

Fig.\ \ref{fig:scalabilityRMSE} shows the average RMSE, i.e., the average of all MA self-localization and object localization errors, 
averaged over all time steps and simulation runs, versus the network size. SPF performs poorly since the numbers of particles it uses are not sufficient 
to properly represent the high-dimensional distributions. UPF performs much better; in fact, for $J \!=\! 5000$, it outperforms all the other methods. 
This is due to a smart selection of the proposal distribution using the unscented transform. However, the RMSE of both SPF and UPF grows with the network size 
(although for UPF with $J \!=\! 5000$, this is hardly visible in Fig.\ \ref{fig:scalabilityRMSE}). In contrast, the RMSE of PM and NBP, which are both based on 
BP, does not depend on the network size. The RMSE of PM is only slightly higher than that of UPF with $J \!=\! 5000$. 
The RMSE of NBP is higher than that of UPF and PM but considerably lower than that of SPF; it is reduced when $J$ is increased from $500$ to $1000$.
In contrast, the RMSE of PM is effectively equal for $J \!=\! 1000$ and $5000$. Thus, one can conclude that the performance of PM with $J \!=\! 1000$ 
cannot be improved upon by increasing $J$, and the (small) performance gap between UPF and PM is caused by the approximate nature of loopy BP. 
Finally, our simulations also showed that the performance gap between the BP-based methods, PM and NBP, and UPF is increased when the driving noise 
is increased and/or the measurement noise is decreased. This is again due to the smart selection of the proposal distribution in UPF.

These results demonstrate specific advantages of PM over particle filtering methods. In particular, PM has a very good performance-complexity 
tradeoff, and its scaling characteristic with respect to the network size is only 
linear.\footnote{In 
a distributed implementation, the computational complexity of the consensus scheme depends on the number of consensus iterations, $C$, and, 
in the ``consensus--over-weights'' case, on the diameter of the communication graph, $I$. Therefore, the scaling might be slightly higher than 
linear.} 
For small networks, this may come at the cost of a slight performance loss relative to UPF (not, however, relative to SPF, which performs much worse). 
For large networks, the performance of PM can be better than that of UPF, since for a fixed number of particles, the performance of UPF 
decreases with increasing network size (in Fig. \ref{fig:scalabilityRMSE}, this is visible for $J=1000$ but only barely for $J=5000$).

An important further advantage of PM applies to distributed scenarios. Contrary to particle filters, PM facilitates a distributed implementation 
since it naturally distributes the computation effort among the agents. This distribution requires only communication with neighbor agents, 
and the communication cost is typically much smaller than for 
distributed particle filters. For example, in the case of the largest simulated network size of $|\tilde{\cl{A}}| + |\cl{O}| \rmv=\rmv 160$,
for $J \!=\! 1000$, each agent broadcasts 32000 real values per consensus iteration to its neighbors. In 
a distributed implementation of UPF, for proposal adaptation alone, each agent has to broadcast to its neighbors
a covariance matrix of size $640\times640$ or, equivalently, 205120 real values per consensus iteration. Additional
communication is required for other tasks, depending on the specific distributed particle filtering algorithm used \cite{hlinkaMag13}. 
Furthermore, for the considered joint CS--DT problem, UPF is ill-suited to large networks also because it involves the inversion and 
Cholesky decomposition of matrices whose dimension grows with the network size. 
\pagebreak 
In large networks, this may lead to numerical problems 
on processing units with limited dynamic range.

\section{Conclusion}
\label{sec:concl}

We proposed a Bayesian framework and methodology for distributed sequential localization of cooperative agents and noncooperative objects 
in mobile networks, based on recurrent measurements between agents and objects and between different agents.
Our work provides a consistent combination of cooperative self-localization (CS) and distributed object tracking (DT) for multiple mobile or static 
agents and objects. Starting from a factor graph formulation of the joint CS--DT problem, we developed a particle-based, distributed belief 
propagation (BP) message passing algorithm. This algorithm employs a consensus scheme for a distributed calculation of the product of the 
object messages. The proposed integration of consensus in particle-based BP solves the problem of accommodating noncooperative network 
nodes in distributed BP implementations. Thus, it may also be useful for other distributed inference problems.

A fundamental advantage of the proposed joint CS--DT method over both separate CS and DT and simultaneous localization and tracking 
(SLAT) is a probabilistic information transfer between CS and DT. This information transfer allows CS to support DT and vice versa. 
Our simulations demonstrated that this principle can result in significant improvements in both agent self-localization and object localization 
performance compared to state-of-the-art methods. Further advantages of our method are its low complexity and its very good scalability 
with respect to the network size. The computation effort is naturally distributed among the agents, using only a moderate amount of 
communication between neighboring agents. We note that the complexity can be reduced further through an improved proposal distribution 
calculation that uses the \emph{sigma point BP} technique introduced in \cite{meyer14sigma}.
Furthermore, the communication requirements can be reduced through the use of parametric representations of messages and beliefs \cite{meyer2013coop}.

The proposed framework and methodology can be extended to accommodate additional tasks (i.e., in addition to CS and DT) 
involving cooperative agents and/or noncooperative objects, such as distributed synchronization \cite{etzlinger14,meyerAsilomar13} 
and cooperative mapping \cite{dedeoglu00}. Another interesting direction for future work is an extension to scenarios involving an unknown 
number of objects \cite{mahler2007statistical,braca13} and object-to-measurement association uncertainty 
\cite{barShalom11,mahler2007statistical,braca13,williams14,meyer15scalable}.

\vspace*{1mm}

\bibliographystyle{ieeetr_noParentheses}
\bibliography{references}

\begin{IEEEbiography}[{\includegraphics[width=25mm,height=32mm,clip,keepaspectratio]{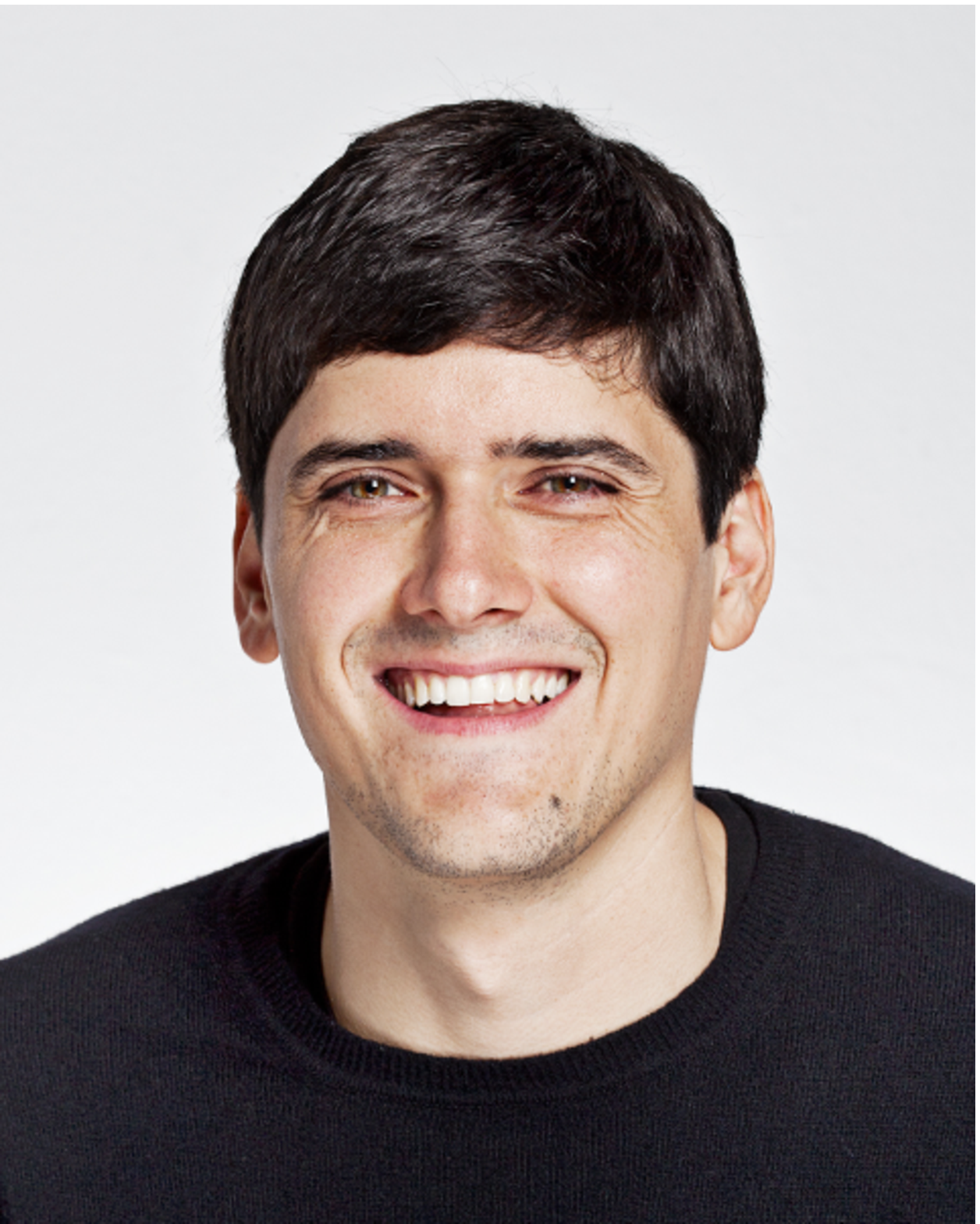}}]{Florian~Meyer} (S'12--M'15) 
received the Dipl.-Ing.\ (M.Sc.)\ and Ph.D.\ degrees in electrical engineering from TU Wien, Vienna, Austria in 2011 and 2015, respectively. Since 2011, he has been a Research and Teaching Assistant with the Institute of Telecommunications, TU Wien. He was a visiting scholar at the Department of Signals and Systems, Chalmers University of Technology, Sweden in 2013 and at the STO Centre of Maritime Research and Experimentation (CMRE), La Spezia, Italy in 2014 and in 2015. His research interests include signal processing for wireless sensor networks, localization and tracking, information-seeking control, message passing algorithms, and finite set statistics.
\end{IEEEbiography}

\begin{IEEEbiography}[{\includegraphics[width=1in,height=1.25in,clip,keepaspectratio]{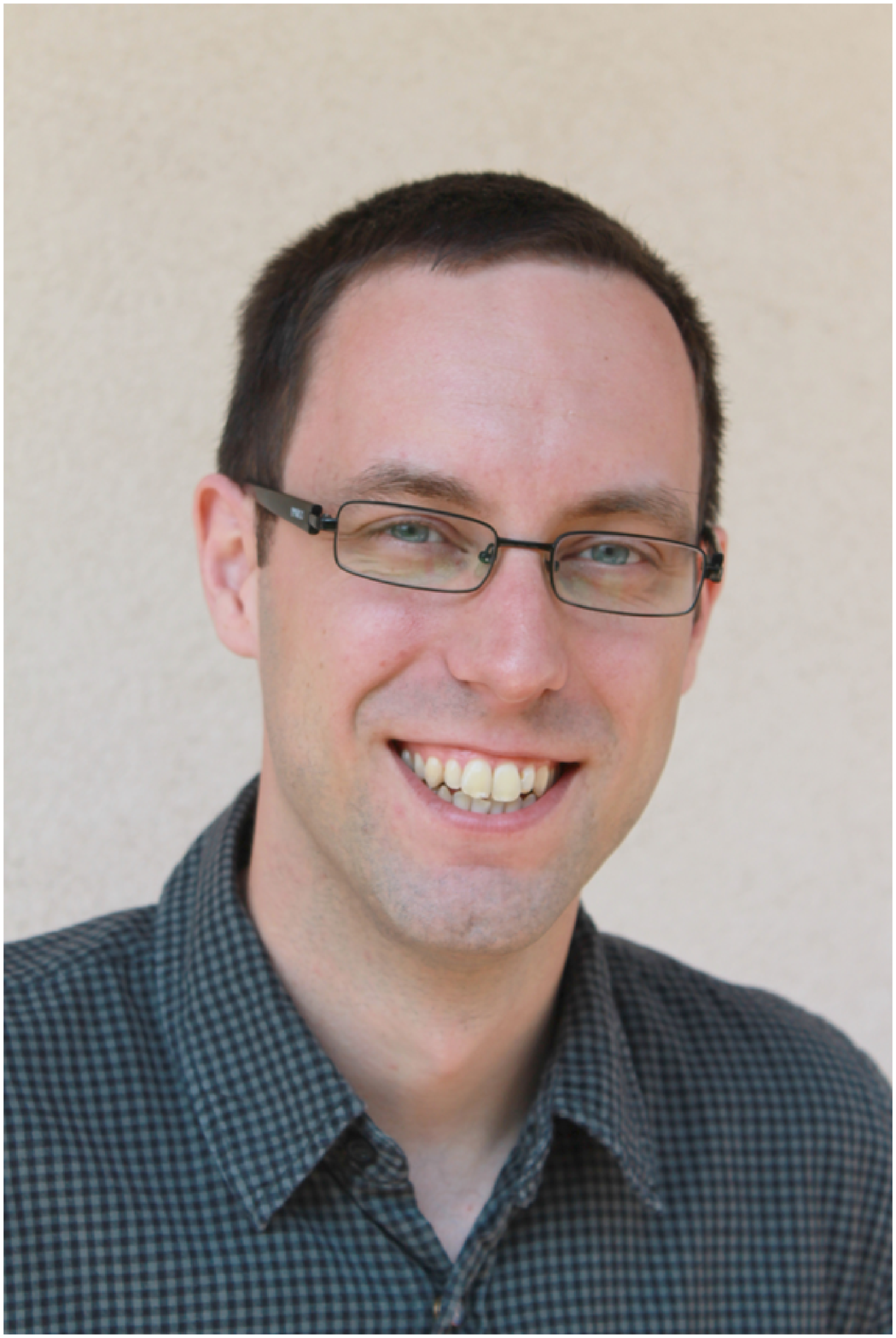}}]{Ondrej Hlinka} 
received the Ing.\ (M.Eng.)\ degree in electrical engineering from the Slovak University of Technology, Bratislava, Slovakia in 2008 and
the Dr.\ techn.\ (Ph.D.)\ degree in electrical engineering/signal processing from TU Wien, Vienna, Austria in 2012. During 2008--2014, he was a research assistant with the Institute of Telecommunications, TU Wien. Since 2014, he has been with robart GmbH, Linz, Austria as a robotics algorithm developer. His research interests include distributed signal processing for agent networks, statistical signal processing, and mobile robot navigation.
\end{IEEEbiography}

\begin{IEEEbiography}[{\includegraphics[width=1in,height=1.25in,clip,keepaspectratio]{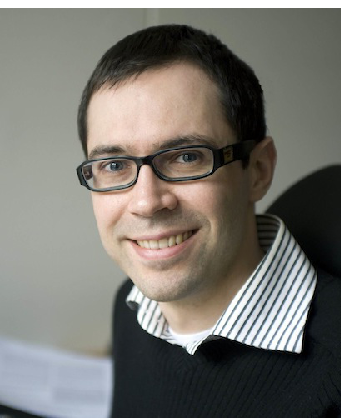}}]{Henk~Wymeersch} (S'99--M'05) received the Ph.D. degree in Electrical Engineering/Applied Sciences in 2005 from Ghent University, Belgium. He is currently an Associate Professor with the Department of Signals and Systems at Chalmers University of Technology, Sweden. Prior to joining Chalmers, he was a Postdoctoral Associate with the Laboratory for Information and Decision Systems (LIDS) at the Massachusetts Institute of Technology (MIT). He served as Associate Editor for \textsc{IEEE Communication Letters} (2009--2013), \textsc{IEEE Transactions on Wireless Communications} (2013--present), and the \textsc{Transactions on Emerging Telecommunications Technologies} (2011--present). 
\end{IEEEbiography}

\begin{IEEEbiography}[{\includegraphics[width=25mm,height=32mm,clip,keepaspectratio]{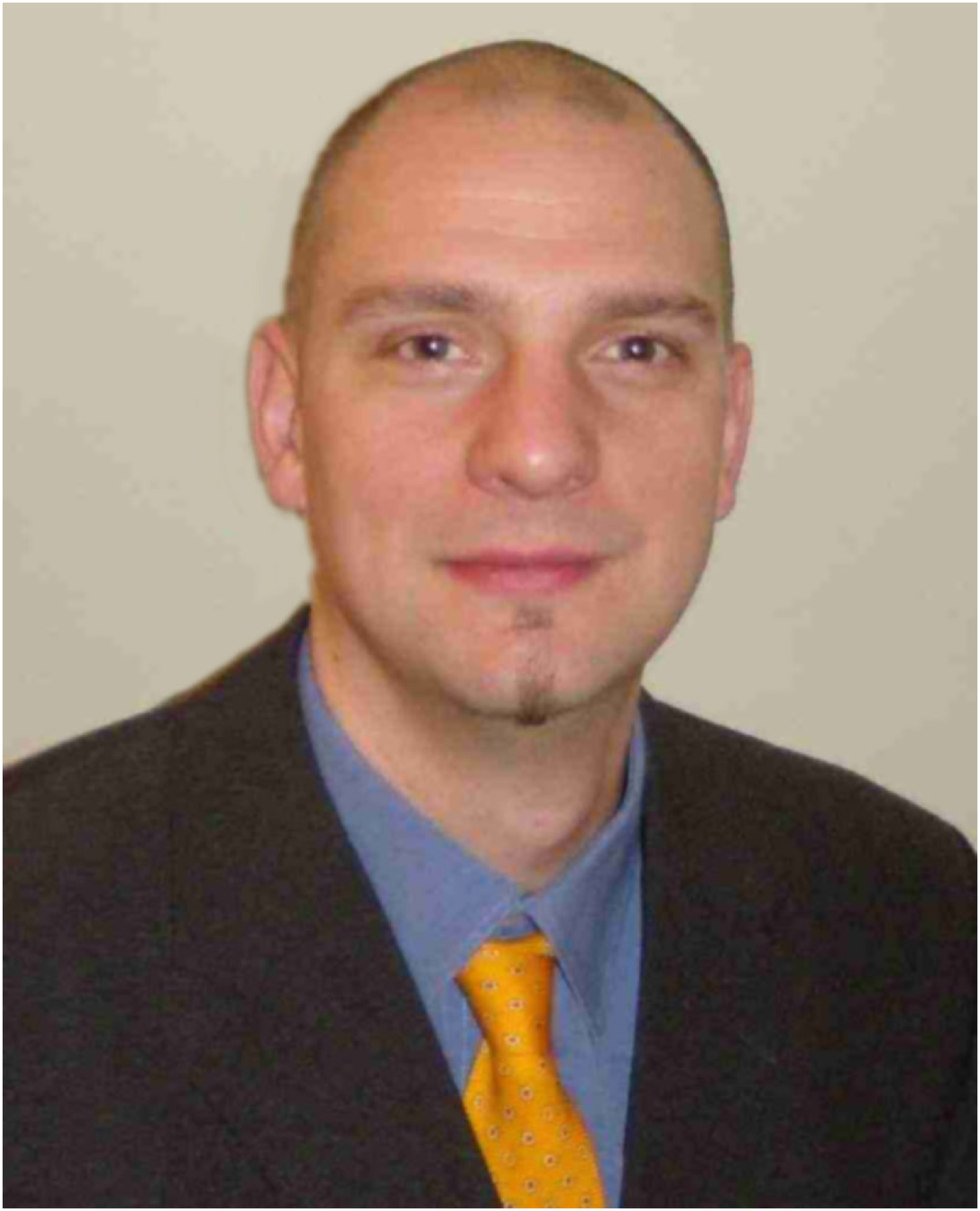}}]{Erwin~Riegler} (M'07) received the Dipl-Ing. degree in Technical Physics (with distinction) in 2001 and the Dr. techn. degree in Technical Physics (with distinction) in 2004 from TU Wien, Vienna, Austria. From 2005 to 2006, he was a postdoctoral researcher at the Institute for Analysis and Scientific Computing, TU Wien. From 2007 to 2010, he was a senior researcher at the Telecommunications Research Center Vienna (FTW). From 2010 to 2014, he was a postdoctoral researcher at the Institute of Telecommunications, TU Wien. Since 2014, he has been a senior researcher at the Swiss Federal Institute of Technology in Zurich (ETHZ). He was a visiting researcher at ETHZ, Chalmers University of Technology, The Ohio State University, Aalborg University, and the Max Planck Institute for Mathematics in the Sciences. His research interests include information theory, noncoherent communications, statistical physics, and transceiver design. He is the co-author of a paper that won a student paper award at the International Symposium on Information Theory, 2012.
\end{IEEEbiography}

\begin{IEEEbiography}[{\includegraphics[width=25mm,height=32mm,clip,keepaspectratio]{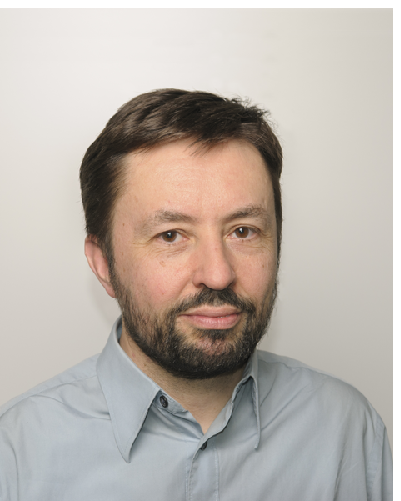}}]{Franz~Hlawatsch}(S'85--M'88--SM'00--F'12) received the Diplom-Ingenieur, Dr. techn., and Univ.-Dozent (habilitation) degrees in electrical engineering/signal processing from TU Wien, Vienna, Austria in 1983, 1988, and 1996, respectively. Since 1983, he has been with the Institute of Telecommunications, TU Wien, where he is currently an Associate Professor. During 1991--1992, as a recipient of an Erwin Schr\"odinger Fellowship, he spent a sabbatical year with the Department of Electrical Engineering, University of Rhode Island, Kingston, RI, USA. In 1999, 2000, and 2001, he held one-month Visiting Professor positions with INP/ENSEEIHT, Toulouse, France and IRCCyN, Nantes, France. He (co)authored a book, three review papers that appeared in the {\sc IEEE Signal Processing Magazine}, about 200 refereed scientific papers and book chapters, and three patents. He coedited three books. His research interests include statistical and compressive signal processing methods and their application to sensor networks and wireless communications.

Prof. Hlawatsch was Technical Program Co-Chair of EUSIPCO 2004 and served on the technical committees of numerous IEEE conferences. He was an Associate Editor for the {\sc IEEE Transactions on Signal Processing} from 2003 to 2007 and for the {\sc IEEE Transactions on Information Theory} from 2008 to 2011. From 2004 to 2009, he was a member of the IEEE SPCOM Technical Committee. He currently serves as an Associate Editor for the {\sc IEEE Transactions on Signal and Information Processing over Networks}. He coauthored papers that won an IEEE Signal Processing Society Young Author Best Paper Award and a Best Student Paper Award at IEEE ICASSP 2011.
\end{IEEEbiography}

\end{document}